\documentclass[10pt,twocolumn,twoside]{IEEEtran}
\usepackage{url}
\usepackage[cmex10]{amsmath}
\usepackage{amsfonts}
\usepackage{amssymb}
\usepackage{pgf}
\usepackage{graphicx}
\usepackage{grffile}
\usepackage{subfig}
\usepackage[font=scriptsize]{caption}
\usepackage{epstopdf}
\usepackage{mathtools}
\usepackage{tabu}
\usepackage{cite}
\newcommand{\ud}{\,\mathrm{d}}
\ifCLASSINFOpdf
\else
\fi
\begin{document}
%
\title{The Fourier Decomposition Method for nonlinear and nonstationary time series analysis}
%
%
%
\author{Pushpendra Singh$^{*,1,3}$, Shiv Dutt Joshi$^1$,
        Rakesh Kumar Patney$^1$,
        and~Kaushik Saha$^2$ 
\thanks{$^*$Corresponding author's e-mail address: spushp@gmail.com; pushpendra.singh@ee.iitd.ernet.in}  
\thanks{$^1$Indian Institute of Technology Delhi, Delhi-110 016, India.}
\thanks{$^2$Samsung R \& D Institute India - Delhi, India.}
\thanks{$^3$Jaypee Institute of Information Technology - Noida, India}}
\markboth{Signal/Data analysis \& processing}
{Singh P. \MakeLowercase{\textit{et al.}}: The Fourier Decomposition Method}
\maketitle
\begin{abstract}
 Since many decades, there is a general perception in literature that the Fourier methods are not suitable for the analysis of nonlinear and nonstationary data. In this paper, we propose a Fourier Decomposition Method (FDM) and demonstrate its efficacy for the analysis of nonlinear (i.e. data generated by nonlinear systems) and nonstationary time series. The proposed FDM decomposes any data into a small number of `Fourier intrinsic band functions' (FIBFs). The FDM presents a generalized Fourier expansion with variable amplitudes and frequencies of a time series by the Fourier method itself. We propose an idea of zero-phase filter bank based multivariate FDM (MFDM) algorithm, for the analysis of multivariate nonlinear and nonstationary time series, from the FDM. We also present an algorithm to obtain cutoff frequencies for MFDM. The MFDM algorithm is generating finite number of band limited multivariate FIBFs (MFIBFs). The MFDM preserves some intrinsic physical properties of the multivariate data, such as scale alignment, trend and instantaneous frequency. The proposed methods produce the results in a time-frequency-energy distribution that reveal the intrinsic structures of a data. Simulations have been carried out and comparison is made with the Empirical Mode Decomposition (EMD) methods in the analysis of various simulated as well as real life time series, and results show that the proposed methods are powerful tools for analyzing and obtaining the time-frequency-energy representation of any data.
\end{abstract}
%
\IEEEpeerreviewmaketitle
\section{Introduction}
\IEEEPARstart{T}{he} time-frequency representation (TFR) is a well established powerful tool for the analysis of time series signals. There exist many types of time-frequency (TF) analysis methods, e.g. linear (the short-time Fourier transform), quadratic (the Wigner-Ville distribution) and Wavelet transforms. The TFR is achieved by formulation often referred as time-frequency distribution (TFD) and provides insight into the complex structure of a signal consisting of several components. 
These approaches have many useful applications, however the analysis of nonstationary signals are not well presented by these methods.

Recently developed Empirical Mode Decomposition (EMD)~\cite{rs1} has provided a general method for examining the TFD, and has been applied to all kinds of data. The EMD is an adaptive signal analysis algorithm for the analysis of nonstationary and nonlinear signals (i.e. signals generated from nonlinear systems). The EMD has become an established method for the signal analysis in various applications, e.g. medical studies~\cite{rs25,rs26,rs30,rs31}, meteorology~\cite{rs1}, geophysical studies~\cite{rs27} and image analysis\cite{rs29}. The EMD decomposes any given data into a finite number of narrow band intrinsic mode functions (IMFs) which are derived directly from the data, whereas other signal decomposition techniques (like the Fourier, Wavelets, etc.) incorporate predefined fixed basis for signal modeling and analysis.
The Ensemble EMD (EEMD) is a noise-assisted data analysis method developed in~\cite{rs4} to overcome the timescale separation problem of EMD. The Multivariate EMD (MEMD) developed in~\cite{rs5} is a generalization of the EMD for multichannel data analysis. The Compact EMD (CEMD) algorithm is proposed in~\cite{rs12} to reduce mode mixing, end effect, and detrend uncertainty present in EMD and to reduce computation complexity of EEMD as well. The IMFs, generated by EMD, are dependent on distribution of local extrema of signal and the type of spline used for upper and lower envelope interpolation. The traditional EMD uses cubic spline for upper and lower envelope interpolation. The EMD algorithm, proposed in~\cite{rs13} to reduce mode mixing and detrend uncertainty, uses nonpolynomial cubic spline interpolation to obtain upper and lower envelopes, and have shown in~\cite{rs14} that it improves orthogonality among IMFs.

The energy preserving property is important for any kind of transformation, and it is obtained by the orthogonal decomposition of signal in various transforms like the Fourier, Wavelet, Fourier-Bessel, etc. The energy preserving property is especially important for the accurate and faithful analysis of three dimensional time-frequency-energy distribution of a signal. The EMD algorithms, proposed in~\cite{rs19}, ensure orthogonality or energy preserving property or both in decomposition of signal into IMFs and refereed to as energy preserving EMD (EPEMD).


In spite of considerable success of EMD, all of the EMD algorithms are based on empirical, heuristic and ad hoc procedure that make them hard to analyze mathematically, and EMD may suffer from mode mixing, detrend uncertainty, aliasing and end effect artefacts~\cite{rs22}. There is a lack of mathematical understanding of the EMD algorithms, e.g. IMFs dependence on the number of sifting and the stopping criteria, convergence property and stability to noise perturbation. In spite of all these limitations, EMD is the widely used nonstationary data analysis method. Therefore, in this paper, EMD is used as a reference to establish the validity, reliability and calibration of the proposed method.

Since many decades, there is an understanding in the literature (e.g.~\cite{rs1,rs22,th4}) that the Fourier methods are, directly, not suitable for nonlinear and nonstationary data analysis, and various reasons (e.g. linearity, periodicity or stationarity) are provided to support it. The Fourier transform is valid under very general Dirichlet conditions (i.e. signal is absolutely integrable with finite number of maxima and minima, and finite number of finite discontinuities in any finite interval) thus include nonlinear and nonstationary signals as well. Therefore, in this study, we explore and provide algorithms to analyze nonlinear and nonstationary data by the Fourier method termed as the Fourier Decomposition Method (FDM), which generates small number of Fourier intrinsic band functions (FIBFs). It is already well established that the Fourier method is best tool for spectrum analysis and, in this study, we show that the Fourier method is also a best tool for time-frequency analysis and processing of any signal. The power of the Fourier transform can also be realized from the fact that the analytic representation and, hence, the Hilbert transform of a signal are, inherently, present in the Fourier transform.


In this paper, we also propose a method, which captures the features of the MEMD, using a zero-phase filter bank (ZPFB) approach to construct the multivariate FIBFs and residue components. This multivariate FDM (MFDM) algorithm generates matched multivariate FIBFs and residue through zero-phase filtering. Thus, we propose an adaptive, data-driven, ZPFB based time-frequency analysis method.

For the adaptive data analysis approach, the most difficult challenge has been to establish a general adaptive decomposition method of analysis without a priori basis. In this study, we propose the FDM and MFDM general adaptive data analysis methods that are inspired by the EMD methods and their filter bank properties~\cite{rs15,rs16}. This paper is organized as follows: In section II the EMD algorithm is briefly presented. We propose the Fourier decomposition method (FDM) in section III. We propose the ZPFB based multivariate FDM algorithm in section IV. Simulation results are presented in section V. Finally conclusions are presented in section VI.
\section{Brief overview of the EMD algorithm}
There are various tools for nonstationary data processing such as the spectrogram; the wavelet analysis; the Wigner-Ville distribution; evolutionary spectrum~\cite{rs28}; the empirical orthogonal function expansion (EOF) (or principal component analysis or singular value decomposition); Synchrosqueezed wavelet transforms~\cite{rs17}; the EMD, etc.

The EMD is a well established signal decomposition method that decomposes nonlinear and nonstationary data into a set of finite band-limited IMFs and residue through the sifting process. The decomposed signal $x(t)$ is expressed as the sum of $\ell$ IMF components plus the final residue as
\begin{equation}
 x(t)=\sum_{i=1}^{\ell}{y}_{i}(t) + r_\ell(t) =\sum_{i=1}^{\ell+1}{y}_{i}(t),\label{emd1}
\end{equation}
where $y_{i}(t)$ is the $i^{th}$ IMF and $r_\ell(t)=y_{\ell+1}(t)$ is final residue.
The IMFs admit amplitude-frequency modulated (AM-FM) representation \cite{rs17} (i.e. $y_i(t) \approx a_i(t)\cos(\phi_i(t))$, with $a_i(t),\frac{\ud\phi_i(t)}{\ud t}=\phi'_i(t)>0$ $\forall t$) and well-behaved Hilbert transforms \cite{rs1}. For any IMF $y_i(t)$, its Hilbert transform $\hat{y_i}(t)$ is defined as convolution of $y_i(t)$ and $1/\pi t$, i.e. $\hat{y_i}(t)=\frac{1}{\pi}\int_{-\infty}^{\infty}\frac{y_i(\tau)}{t-\tau} \ud\tau $ and the Hilbert transform emphasizes the local properties of $y_i(t)$. An analytic signal $z_i(t)$ can be represented by $z_i(t)=y_i(t)+j\hat{y_i}(t)=a_i(t)\exp{(j\phi_i(t))}$, where $a_i(t)=[y_i^2(t)+\hat{y_i}^2(t)]^{1/2}$ and  $\phi_i(t)=\tan^{-1}[\hat{y_i}(t)/y_i(t)]$ are instantaneous amplitude and phase of $y_i(t)$. The instantaneous frequency (IF) of $y_i(t)$ is defined as: $\omega_i(t)=\phi'_i(t)=\frac{\hat{y_i}'(t)y_i(t)-\hat{y_i}(t)y'_i(t)}{\hat{y_i}^2(t)+y_i^2(t)}$. The physical meaning of IF $\omega_i(t)$ constrains that $\phi_i(t)$ must be a mono-component function of time. The Bedrosian and Nuttall theorems \cite{rs20,rs21} further impose non-overlapping spectra constraints on the pair [$a_i(t),\cos(\phi_i(t))$] of a signal $y_i(t)=a_i(t)\cos(\phi_i(t))$.

All IMFs must satisfy two basic conditions: (1) In the complete range of time series, the number of extrema (i.e. maxima and minima) and the number of zero crossings are equal or differ at most by one. (2) At any point of time in the complete range of time series, the average of the values of upper and lower envelopes, obtained by the interpolation of local maxima and the local minima, is zero. The first condition ensure that IMFs are narrow band signals and the second condition is necessary to ensure that the IF does not have redundant fluctuations because of asymmetric waveforms~\cite{rs1}.


\section{The Fourier Decomposition Method}
We propose a class of functions, termed as the Fourier intrinsic band functions (FIBFs), belonging to $C^{\infty}[a,b]$, here with the following formal
definition.
\newtheorem{definition}{Definition}
\begin{definition} \label{law:box}
The Fourier intrinsic band functions (FIBFs), $y_i(t) \in C^{\infty}[a,b]$, are functions that satisfy the following conditions:
\\ (1) The FIBFs are zero mean functions, i.e. $\int_a^b y_i(t)\ud t=0$.
\\ (2) The FIBFs are orthogonal functions, i.e. $\int_a^b y_i(t) y_j(t)\ud t=0$, for $i\ne j$.
\\ (3) The FIBFs provide analytic FIBFs (AFIBFs) with instantaneous frequency (IF) and amplitude always greater than zero, i.e. $y_i(t)+j\hat{y_i}(t)=a_i(t)\exp(j\phi_i(t))$, with $a_i(t), \frac{\ud}{\ud t} \phi_i(t) \ge0$, $\forall t$.
\end{definition}
 Thus, the AFIBFs are monocomponent signals and, physically, the IF has meaning only for monocomponent signals, i.e., signal has only one frequency or a narrow range of frequencies varying as a function of time~\cite{th4}. Thus, the FIBF is sum of zero mean sinusoidal functions of consecutive frequency band.

The main objective of this study is to obtain unique representation of multicomponent signal as a sum of constant and monocomponent signals, i.e. signals which can be represented by the following model~\cite{th4}:
\begin{equation}
x(t)= \sum_{i=1}^{M}y_i(t)+n(t), \label{EqObj}
\end{equation}
where $n(t)$ is a noise representing any residue (constant or trend) components and the $y_i(t)$ are $M$ single component nonstationary signals, which in our proposed framework would be the FIBFs, defined above.

The necessary conditions~\cite{rs1}, for a basis to represent a nonlinear and nonstationary time series, are completeness, orthogonality, locality, and adaptiveness. The FIBFs, intrinsically, follow all the necessary conditions by virtue of the decomposition.

The available data are usually of finite duration, nonstationary and generated from the systems that are generally nonlinear.
Let $x(t)$ be a time limited $[t_1, t_1+T_0]$ real valued signal which follows the Dirichlet conditions.
We construct the periodic signal as $x_{\text{\tiny T}_0}(t)=\sum_{s=-\infty}^{\infty} x(t-sT_0)$ such that $x(t)={x_{\text{\tiny T}_0}(t)}w(t)$, where $w(t)=1$, for $t_1\le t\le t_1+T_0$ and zero otherwise. The Fourier series expansion of $x_{\text{\tiny T}_0}(t)$ is given by
\begin{equation}
    x_{\text{\tiny T}_0}(t)=a_0+\sum_{k=1}^{\infty} [a_k\cos(k\omega_0 t)+b_k\sin(k\omega_0 t)], \label{eq1}
\end{equation}
where frequency (rad/s) $\omega_0=\frac{2\pi}{T_0}$, $a_0=\frac{1}{T_0}\int_{t_1}^{t_1+T_0} x_{\text{\tiny T}_0}(t)\ud t$, $a_k=\frac{2}{T_0}\int_{t_1}^{t_1+T_0} x_{\text{\tiny T}_0}(t)\cos(k\omega_0 t)\ud t$ and $b_k=\frac{2}{T_0}\int_{t_1}^{t_1+T_0} x_{\text{\tiny T}_0}(t)\sin(k\omega_0 t)\ud t$.
We write \eqref{eq1} as
\begin{equation}
x_{\text{\tiny T}_0}(t)=a_0+\frac{1}{2}\sum_{k=1}^{\infty}[c_k \exp{(jk\omega_0t)}+c^*_k \exp{(-jk\omega_0t)}], \label{eq2}
\end{equation}
where $c_k=(a_k-jb_k)$ and $c^*_k=(a_k+jb_k)$. From \eqref{eq2}, it is clear that
\begin{equation}
x_{\text{\tiny T}_0}(t)=a_0+Re\{z_{\text{\tiny T}_{0}}(t)\}, \label{eq3}
\end{equation}
where analytic function
\begin{equation}
z_{\text{\tiny T}_{0}}(t) \triangleq \sum_{k=1}^{\infty}c_k \exp{(jk\omega_0t)} \label{eq31}
\end{equation}
is complex conjugate of $\tilde{z}_{\text{\tiny T}_{0}}(t) \triangleq \sum_{k=1}^{\infty}c^*_k \exp{(-jk\omega_0t)}$ and $Re\{z_{\text{\tiny T}_{0}}(t)\}$ denotes the real part of $z_{\text{\tiny T}_{0}}(t)$.
We write $z_{\text{\tiny T}_{0}}(t)$ as
 \begin{equation}
 z_{\text{\tiny T}_{0}}(t)=\sum_{i=1}^{M}a_i(t) \exp{(j\phi_i(t))}, \label{eq41}
 \end{equation}
 where, in forward search (i.e. low to high frequency scan) of AFIBFs, $a_1(t)\exp{(j\phi_1(t))}=\sum_{k=1}^{N_1}c_k \exp{(jk\omega_0t)}$, $a_2(t)\exp{(j\phi_2(t))}=\sum_{k=(N_1+1)}^{N_2}c_k \exp{(jk\omega_0t)}$, $\cdots$, $a_M(t)\exp{(j\phi_M(t))}=\sum_{k=(N_{M-1}+1)}^{\infty}c_k \exp{(jk\omega_0t)}$, i.e.
\begin{equation}
a_i(t)\exp{(j\phi_i(t))}=\sum_{k=N_{i-1}+1}^{N_i}c_k \exp{(jk\omega_0t)}, \label{eq42}
\end{equation}
with $N_0=0$ and $N_M=\infty$. The FIBFs are the real part of AFIBFs presented in Eq.~\eqref{eq42}. In order to obtain minimum number of AFIBFs in low to high frequency scan (LTH-FS), for each $i$, start with $(N_{i-1}+1)$ and append more term till we reach the maximum value of $N_i$ such that $(N_{i-1}+1) \le N_i \le \infty$ and
\begin{equation}
a_i(t),\omega_i(t)=\frac{\ud\phi_i(t)}{\ud t}\ge0, \forall t. \label{eq42v1}
\end{equation}
It is easy to observe that such a decomposition is always possible.

Similarly, in reverse search (i.e. form high to low frequency scan (HTL-FS)) of AFIBFs, we obtain $a_1(t)\exp{(j\phi_1(t))}=\sum_{k=N_1}^{\infty}c_k \exp{(jk\omega_0t)}$, $a_2(t)\exp{(j\phi_2(t))}=\sum_{k=N_2}^{(N_1-1)}c_k \exp{(jk\omega_0t)}$, $\cdots$, $a_M(t)\exp{(j\phi_M(t))}=\sum_{k=1}^{(N_{M-1}-1)}c_k \exp{(jk\omega_0t)}$, and the lower and upper limits of sum in Eq.~\eqref{eq42} would change to $k=N_{i} \text{ to } (N_{i-1}-1)$, respectively, with $N_0=\infty$, $N_M=1$. Here, we start with $(N_{i-1}-1)$, decrease and select minimum value of $N_i$ such that $1 \le N_i \le (N_{i-1}-1)$ and Eq.~\eqref{eq42v1} is satisfied for $i=1,\cdots,M$.

Observe that ~\eqref{eq41} has precisely the form that in the literature~\cite{rs1} is termed as a generalized Fourier expansion. Moreover, it is worth noting that this representation is complete, orthogonal, local, adaptive and purely Fourier based.
Thus, we have obtained a generalized Fourier expansion of a time series in Eq.~\eqref{eq41} by the Fourier method itself. The variable amplitude and the IF have improved the efficiency of the expansion by expanding the signal into finite number of analytic FIBFs, in \eqref{eq41}, and enabled the expansion to accommodate nonstationary data. Thus, we have obtained a variable amplitude and frequency representation, whereas, the classical Fourier expansion provides the constant amplitude and fixed-frequency representation.

For each FIBFs, the amplitude $a_i(t)$ and IF $f_i(t)$ are functions of time, therefore, we define the three dimensional $\{t,f_i(t),a_i(t)\}$ time-frequency distribution of amplitude as the Fourier-Hilbert spectrum (FHS) $H(f,t)$. The marginal Hilbert spectrum (MHS), derived from Hilbert spectrum, is defined in~\cite{rs1}. Similarly, here we define the marginal Fourier-Hilbert spectrum (MFHS) from the FHS as follows:
\begin{equation}
h(f)=\int_{0}^{T_0}H(f,t)\ud t. \label{FDM_eq7}
\end{equation}
The marginal Fourier-Hilbert spectrum offers a measure of total amplitude (or energy) contribution from each value of frequency in a probabilistic sense. The frequency in either $H(f, t)$ or $h(f)$ has a different meaning from the Fourier spectral analysis~\cite{rs1}. The presence of energy at each frequency in MFHS $h(f)$ means that, in the total duration of the signal, there is a higher likelihood for such a wave (FIBF) to have appeared locally. The frequency in the MFHS indicates only the likelihood that an oscillation with such a frequency exists. The exact occurrence time of that oscillation is given in the full Fourier-Hilbert spectrum. We can also define the instantaneous energy density, which can be used to measure the fluctuation of energy with time, as
\begin{equation}
E(t)=\int_{0}^{f_M}H^2(f,t)\ud f, \label{FDM_eq8}
\end{equation}
where $f_M$ is a maximum frequency of signal.
From Eq.~\eqref{eq1}, we obtain the energy of signal $x(t)$ (or power of signal $x_{\text{\tiny T}_0}(t)$) by the Parseval's theorem as $E_x=a^2_0+\frac{1}{2}\sum_{i=1}^\infty[a^2_n+b^2_n]$ and from Eq.~\eqref{eq31} energy of the analytic signal (or power of signal $z_{\text{\tiny T}_{0}}(t)$) as $E_z=\sum_{i=1}^\infty[a^2_n+b^2_n]$, therefore, relation between $E_x$ and $E_z$ is given by
\begin{equation}
E_x=a^2_0+\frac{E_z}{2}. \label{ExEz}
\end{equation}
Hence, energy of zero mean signal is half of the energy of its analytic signal.

Since in practice the continuous time signals are, generally, discretized for further processing by a computing device, so we present the FDM for discrete signal. Let, $x[n]$, be a discrete signal of length $N$. Using the discrete Fourier transform (DFT), we can write $x[n]$ as
\begin{equation}
    x[n]=\sum_{k=0}^{N-1}X[k]\exp(\frac{j2\pi kn}{N}), \label{eq4}
\end{equation}
where $X[k]=\frac{1}{N}\sum_{n=0}^{N-1}x[n]\exp(\frac{-j2\pi kn}{N})$ is the DFT of signal $x[n]$. 
Let $N$ be an even number (we can proceed in the similar fashion when $N$ is an odd number), then $X[0]$ and $X[\frac{N}{2}]$ are real numbers; and we can write $x[n]$ as
\begin{multline}
    x[n]=X[0]+\sum_{k=1}^{\frac{N}{2}-1}X[k]\exp(\frac{j2\pi kn}{N}) + X[\frac{N}{2}]\exp(j\pi n)\\+ \sum_{k=\frac{N}{2}+1}^{N-1}X[k]\exp(\frac{j2\pi kn}{N}).\label{eq6}
\end{multline}
Since $x[n]$ is real, therefore, $z_1[n] \triangleq \sum_{k=1}^{\frac{N}{2}-1}X[k]\exp(\frac{j2\pi kn}{N})$ is complex conjugate of $z_2[n] \triangleq \sum_{k=\frac{N}{2}+1}^{N-1}X[k]\exp(\frac{j2\pi kn}{N})$ and we can write \eqref{eq6} as
\begin{equation}
    x[n]=X[0]+2Re\{z_1[n]\} + X[\frac{N}{2}](-1)^n, \label{eq7}
\end{equation}
where $Re\{z_1[n]\}$ denote the real part of $z_1[n]$. Now, we write analytic signal $z_1[n]$ as
\begin{equation}
\sum_{k=1}^{\frac{N}{2}-1}X[k]\exp(\frac{j2\pi kn}{N})=\sum_{i=1}^{M}a_i[n]\exp(j\phi_i[n]), \label{eq8}
\end{equation}
where, in forward search (i.e. low to high frequency scan) of AFIBFs, we obtain $a_1[n]\exp{(j\phi_1[n])}=\sum_{k=1}^{N_1}X[k]\exp(\frac{j2\pi kn}{N})$, $a_2[n]\exp{(j\phi_2[n])}=\sum_{k=(N_1+1)}^{N_2}X[k]\exp(\frac{j2\pi kn}{N})$, $\cdots$, $a_M[n]\exp{(j\phi_M[n])}=\sum_{k=(N_{M-1}+1)}^{\frac{N}{2}-1}X[k]\exp(\frac{j2\pi kn}{N})$, i.e.
\begin{equation}
a_i[n]\exp(j\phi_i[n]=\sum_{k=N_{i-1}+1}^{N_{i}}X[k]\exp(\frac{j2\pi kn}{N}), \label{eq9}
\end{equation}
with $N_0=0$ and $N_M=(\frac{N}{2}-1)$.
In order to obtain minimum number of FIBFs in LTH-FS, for each $i$, we scan from ($N_{i-1}+1$) to $(\frac{N}{2}-1)$, obtain maximum value of $N_i$ such that $(N_{i-1}+1) \le N_i \le (\frac{N}{2}-1)$ and phase $\phi_i[n]$ is a monotonically increasing function, i.e. $\omega_i[n]=(\phi_i[n+1]-\phi_i[n])\ge0$ or
\begin{equation}
\omega_i[n]=(\frac{\phi_i[n+1]-\phi_i[n-1]}{2})\ge0, \quad \forall n \label{monotoneeq9}
\end{equation}
and $a_i[n]\ge0$ for $i=1,\cdots,M$ and $\forall n$.
Observe that such a decomposition always exists.

Similarly, in HTL-FS for FIBFs, we obtain $a_1[n]\exp{(j\phi_1[n])}=\sum_{k=N_1}^{\frac{N}{2}-1}X[k]\exp(\frac{j2\pi kn}{N})$, $a_2[n]\exp{(j\phi_2[n])}=\sum_{k=N_2}^{N_1-1}X[k]\exp(\frac{j2\pi kn}{N})$, $\cdots$, $a_M[n]\exp{(j\phi_M[n])}=\sum_{k=1}^{(N_{M-1}-1)}X[k]\exp(\frac{j2\pi kn}{N})$, and the lower and upper limits of sum in Eq.~\eqref{eq9} will change to $k=N_{i} \text{ to } (N_{i-1}-1)$, respectively, with $N_0=\frac{N}{2}$, $N_M=1$. Here, for each $i$, we scan from ($N_{i-1}-1$) to $1$, obtain the minimum value of $N_i$ such that $1 \le N_i \le (N_{i-1}-1)$ and phase $\phi_i[n]$ is a monotonically increasing function.

Thus, FDM provides two views, low to high frequency and high to low frequency view, of the signal and generate two set of time-frequency-energy distribution. Depending on the signal, both view may be same or sometimes they reveal two different types of features of the signal. The FDM is summarized in Algorithms A and B. The FDM with Fourier transform (FT) and discrete time Fourier transform (DTFT) is summarized in appendix.

\noindent\begin{tabular}{p{0.47\textwidth}}
\hline
Algorithm A: The FDM algorithm (LTH-FS) to obtain AFIBFs, for $i=1,\cdots,M$ with $N_0=0$ and $N_M=\frac{N}{2}.$\\
\hline
${\text{STEP }1.}$ Obtain $X[k]=FFT\{x[n]\}$.\\
${\text{STEP }2.}$ Set $AFIBF_i=\sum_{k=(N_{i-1}+1)}^{N_{i}}X[k]\exp(\frac{j2\pi kn}{N})=a_i[n]\exp(j\phi_i[n])$, obtain maximum value of $N_i$ such that $(N_{i-1}+1) \le N_i \le (\frac{N}{2}-1)$ and phase $\phi_i[n]$ of $AFIBF_i$ is a monotonically increasing function, i.e. $\omega_i[n]=(\frac{\phi_i[n+1]-\phi_i[n-1]}{2})\ge0$, $\forall n$. \\
\hline
\end{tabular}

\noindent\begin{tabular}{p{0.47\textwidth}}
\hline
Algorithm B: The FDM algorithm (HTL-FS) to obtain AFIBFs, for $i=1,\cdots,M$ with $N_0=\frac{N}{2}$ and $N_M=1.$\\
\hline
${\text{STEP }1.}$ Obtain $X[k]=FFT\{x[n]\}$.\\
${\text{STEP }2.}$ Set $AFIBF_i=\sum_{k=N_{i}}^{(N_{i-1}-1)}X[k]\exp(\frac{j2\pi kn}{N})=a_i[n]\exp(j\phi_i[n])$, obtain minimum value of $N_i$ such that $1 \le N_i \le (N_{i-1}-1)$ and phase $\phi_i[n]$ of $AFIBF_i$ is a monotonically increasing function, i.e. $\omega_i[n]=(\frac{\phi_i[n+1]-\phi_i[n-1]}{2})\ge0$, $\forall n$.\\
\hline
\end{tabular}
\section{Multivariate Fourier Decomposition Method}
From~\eqref{eq42} and~\eqref{eq9}, we observe that the operation that generates the FIBFs is nothing but the Fourier based zero-phase filtering (ZPF). This is another motivation, in addition to the FB properties of IMFs, to use the Fourier or other methods of zero-phase filtering to decompose any data into a set of FIBFs.
The ZPF of a real valued signal $x[n]$ by zero-phase filter ($h_{z}[n]$) can be obtained by two methods: (1) Convolution method, i.e. $y[n]=x[n]*(h_{z}[n])\Rightarrow Y[k]=X[k]H_{z}[k]$, where $h_{z}[n]=h[n]*h[-n]$ and $H_{z}[k]=|H[k]|^2$, where $h[n]$ is a real sequence. (2) The Fourier method, i.e. set $H_{z}[k]=1$ at desired frequency band and $H_{z}[k]=0$ otherwise, obtain output by the inverse DFT, i.e. $y[n]=\sum_{k=0}^{N-1}X[k]H_{z}[k]\exp(j2\pi kn/N)$, where $X[k]=\frac{1}{N}\sum_{k=0}^{N-1}x[n]\exp(-j2\pi kn/N)$.

We use ZPF that does not shift the essential features of the signal, and propose a multivariate FDM (MFDM) algorithm to generate multivariate FIBFs (MFIBFs) and residue as follows: Apply zero-phase high pass filtering (ZP-HPF) with cutoff frequency $f_{c1}$ to each of the components of the P-variate (P-channel) time series $\{x_p(t)\}_{p=1}^{P}$ and obtain first set of MFIBF ${y_{p1}(t)}$. The first set of residue is obtained as follows:
\begin{equation}
{r_{p1}(t)=x_p(t)-y_{p1}(t)}, \qquad p=1,2,\cdots, P.\label{fbemd1}
\end{equation}
 Apply ZP-HPF with cutoff frequency $f_{c2}$ to set of residue ${r_{p1}(t)}$ and obtain second set of MFIBF ${y_{p2}(t)}$. The second set of residue is obtained as
 \begin{equation}
 {r_{p2}(t)=r_{p1}(t)-y_{p2}(t)} \qquad p=1,2,\cdots,P.\label{fbemd2}
 \end{equation}
We can repeat this ZP-HPF procedure $\ell$ times and obtain final set of MFIBF ${y_{p\ell}(t)}$ and residue (with cutoff frequency $f_{c\ell}$)
\begin{equation}
{r_{p\ell}(t)=r_{p(\ell-1)}(t)-y_{p\ell}(t)} \qquad p=1,2,\cdots,P.\label{fbemd3}
\end{equation}
Through the addition of \eqref{fbemd1}, \eqref{fbemd2} and \eqref{fbemd3} we obtain expression, similar to \eqref{emd1}, for P-variate time series as
\begin{equation}
 x_p(t)=\sum_{i=1}^{\ell}{y}_{pi}(t) + r_{p\ell}(t),\qquad p=1,2,\cdots,P. \label{fbemd4}
\end{equation}
 When we use the Fourier based zero-phase filtering, as in Eq.~\eqref{eq42}, to obtain MFIBFs, first two conditions of FIBFs are fully satisfied and the third one is approximately satisfied (i.e. satisfied in all practical sense), obviously, it can not be guaranteed simultaneously to all P-channel data. This is similar to MEMD algorithm problem where in derivation of multivariate IMFs, first condition of IMF is not imposed~\cite{rs5}.

The question is, how to obtain cutoff frequencies (CFs) $f_{c1},f_{c2},\cdots,f_{c\ell}$ corresponding to zero-phase high pass filters $h_{z1}(t),h_{z2}(t),\cdots,h_{z\ell}(t)$? There is lot of flexibility and are various ways to select CFs, e.g. dyadic (i.e. $f_{c1}=\frac{f_{M}}{2},f_{c2}=\frac{f_{M}}{2^2},\cdots,f_{c\ell}=\frac{f_{M}}{2^\ell}$, where $f_{M}$ is the maximum frequency of a signal $x(t)$ and for the sampled signal, maximum frequency is ($\frac{F_s}{2}$) half of the sampling frequency), non-dyadic, uniform and non-uniform CFs. We can take the Fourier transform of signal $x(t)$ to obtain its spectrum details and make strategy to decide CFs.

For narrowband signal, we define ratio of center frequency ($f_C$) to bandwidth (BW) as
\begin{equation}
m=f_{Ci}/(f_{Hi}-f_{Li}),\quad f_{Ci}=(f_{Hi}+f_{Li})/2, \label{CF1}
\end{equation}
 where $f_{Hi}$ is the highest frequency and $f_{Li}$ is the lowest frequency of $i$th band of a filter bank. From \eqref{CF1} we obtain
\begin{equation}
f_{Li}=[(2m-1)/(2m+1)]f_{Hi},\quad m>1/2. \label{CF2}
\end{equation}
From \eqref{CF1} and \eqref{CF2}, we observe that the ratios, for the consecutive $i$th and  $(i+1)$th bands, of center frequencies $f_{Ci}$, CFs $f_{ci}$ and BWs $(f_{Hi}-f_{Li})$ can be taken as a constant, i.e.
\begin{equation}
f_{Ci}/f_{Ci+1}=f_{ci}/f_{ci+1}=\frac{(f_{Hi}-f_{Li})}{(f_{Hi+1}-f_{Li+1})}=l.\label{CF3}
\end{equation}
From \eqref{CF1}, \eqref{CF2} and \eqref{CF3}, we obtain $l=(2m+1)/(2m-1)$ or $m=(1/2)(l+1)/(l-1)$ with $l>1$, and as [${m \to \infty}$, ${l \to 1}$], [${m \to 1/2}$, ${l \to \infty}$]. Here, we have liberty to select any suitable value of $l$ or $m$, and greater the value of $m$ (or lesser the value of $l$) narrower the band, whereas in the case of dyadic FB $l=2$ and $m=1.5$ are fixed values. If required, we can vary the value of $m$ (or $l$) for each band rather than taking the fixed value. Thus, we here propose the compact and elegant way to decide CFs as summarized in Algorithm C.
\begin{tabular}{p{0.47\textwidth}}
\hline
Algorithm C: An algorithm to obtain cutoff frequencies $f_{ci}$.\\
\hline
${1.}$ Select suitable value of $m$ and set $f_{H1}=F_s/2$.
\\${2.}$ Set $f_{ci}=[(2m-1)/(2m+1)]f_{Hi}$.
\\${3.}$ Set $f_{Hi+1}=f_{ci}$.
\\${4.}$ Repeat step 2 to 3 for $i=1,2,\cdots,\ell$.\\
\hline
\end{tabular}

In MFDM, we can use zero-phase low pass filtering (ZP-LPF) in place of ZP-HPF to decompose signal in order of residue to first MFIBFs, i.e. $r_{p\ell}(y), y_{p\ell}(t), \cdots,y_{p1}(t)$. We use zero-phase filtering as it preserves salient features (e.g. maxima, minima, etc.) in the filtered time waveform exactly at the time where those features occur in the unfiltered waveform, whereas conventional (non zero-phase) filtering shifts the features in the signal and hence cannot be used.
The zero-phase filtering of time series can be obtained through the finite impulse response (FIR) or infinite impulse response (IIR) filters.

Similar to the MEMD and noise-assisted MEMD (NA-MEMD)~\cite{rs16}, this MFDM algorithm produces the equal number of scale-aligned MFIBFs for all channels and preserving joint channel properties that make it suitable for direct multichannel modelling. The FDM does not suffer from mode mixing, detrend uncertainty and end effect artefacts as extraction of FIBFs does not depend on distribution of local extrema across the range of signal.

\begin{table*}[ht]
\caption{Comparisons among the Fourier, Wavelet, EMD-Hilbert and proposed FDM methods in data analysis.} 
\centering 
\begin{tabular}{|c | c| c| c |c |} 
\hline 
  & Fourier  & Wavelet  & EMD-Hilbert & FDM \\ [0.5ex] 
\hline
Basis & a priori & a priori & adaptive & a priori $\mapsto$ adaptive \\\hline   
Frequency & convolution: & convolution: & differentiation: & differentiation:  \\  
 & global & regional & local & local  \\\hline  
Uncertainty & yes & yes & no &  no  \\\hline 
Presentation & frequency- & frequency- & frequency- & frequency-  \\  
 & energy & time-energy & time-energy & time-energy   \\\hline  
Nonlinear & no & no & yes & yes   \\\hline 
Nonstationary & no & yes & yes &   yes  \\\hline 
Harmonics & yes & yes & no & yes $\mapsto$ no  \\\hline 
Theoretical & complete & complete & empirical &  complete  \\ 
base &  &  &  &    \\ 
\hline 
\end{tabular}
\label{table:ComFWEF} 
\end{table*}

Table~\ref{table:ComFWEF} presents comparisons among Fourier, Wavelet, EMD~\cite{CompFHT} and FDM methods in data analysis.
\section{Simulation results}
The online available MATLAB software of MEMD~\cite{rs23}, EMD and EEMD~\cite{rs24} have been used in simulation results.
\subsection{Multivariate data decomposition}
\begin{figure}[!t]
\centering
\includegraphics[angle=0,width=0.5\textwidth,height=0.3\textwidth]{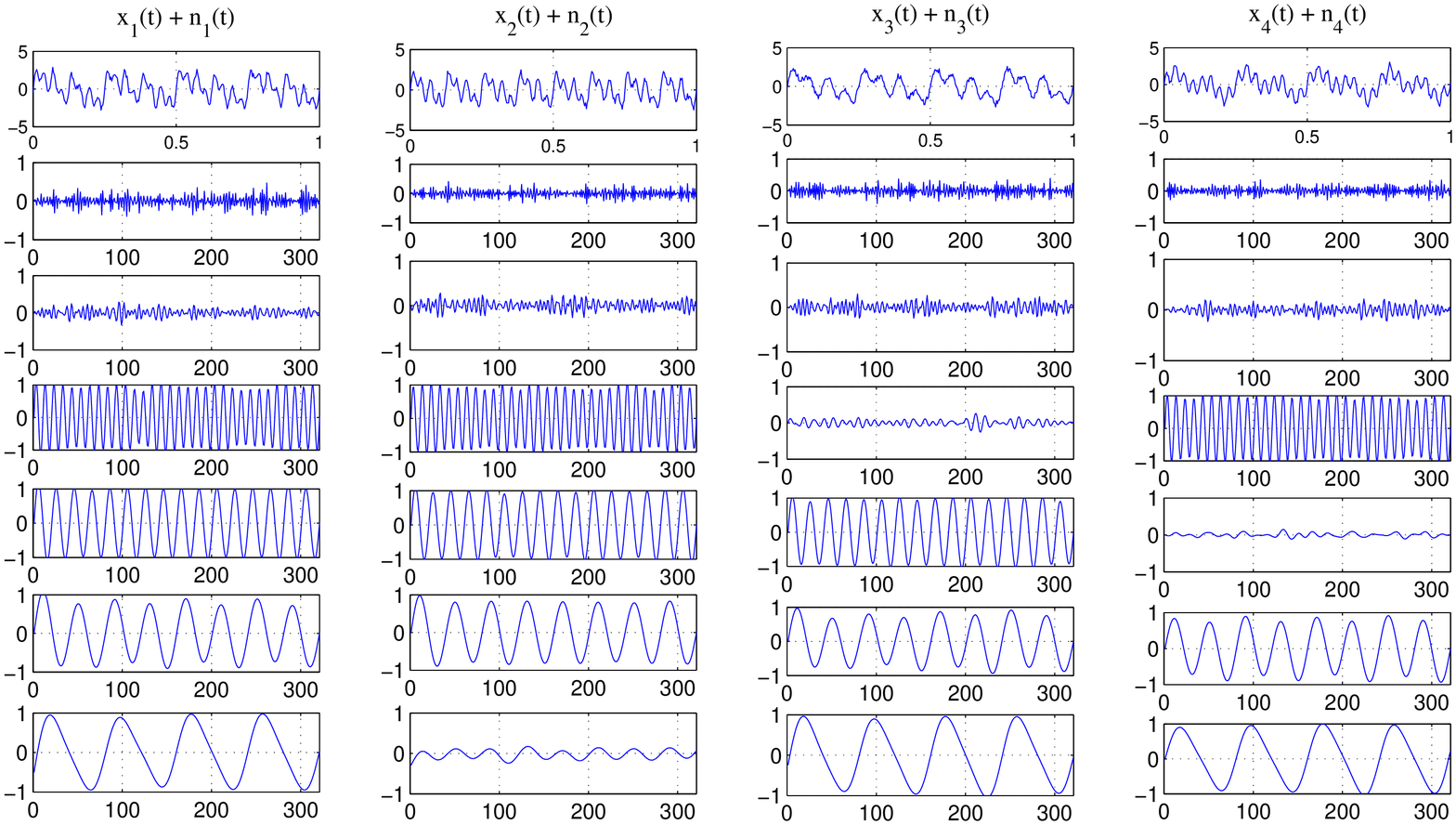}
\captionof{figure}{MFDM applied to a quadri-variate tone-noise mixture, generating perfectly aligned intrinsic modes in all the four channels.}
\label{fig:MIMFsFB}
\end{figure}
We used quadri-variate time series signal, which is summation of sinusoids (with combination of frequencies $f_1 =4 Hz, f_2 =8 Hz, f_3 =16 Hz, f_4 =32 Hz$) and Gaussian white noise of zero mean and standard deviation of 0.2., i.e. $x_j(t)=\sum_{i\in[1,4]}sin(2\pi f_{i}t)+n_j(t)$ (for $j=1,\cdots,4$). A $32Hz$ sinusoid is present to first, second and fourth channels; a $16Hz$ sinusoid is present to first, second and third channels; a $8Hz$ sinusoid is common to all channels; a $4Hz$ sinusoid is present to first, third and fourth channels.
On the same machine, computation time for MFDM is 0.45 sec. and for MEMD is 69.5 sec. in this simulation. The MFDM algorithm, similar to MEMD, generating perfectly aligned intrinsic bands, as shown in Figure~\ref{fig:MIMFsFB}, in all the four channels, whereas, MFDM is computationally more efficient.
\subsection{Intermittency and mode mixing}
\begin{figure}[!t]
\centering
\includegraphics[angle=0,width=0.5\textwidth,height=0.3\textwidth]{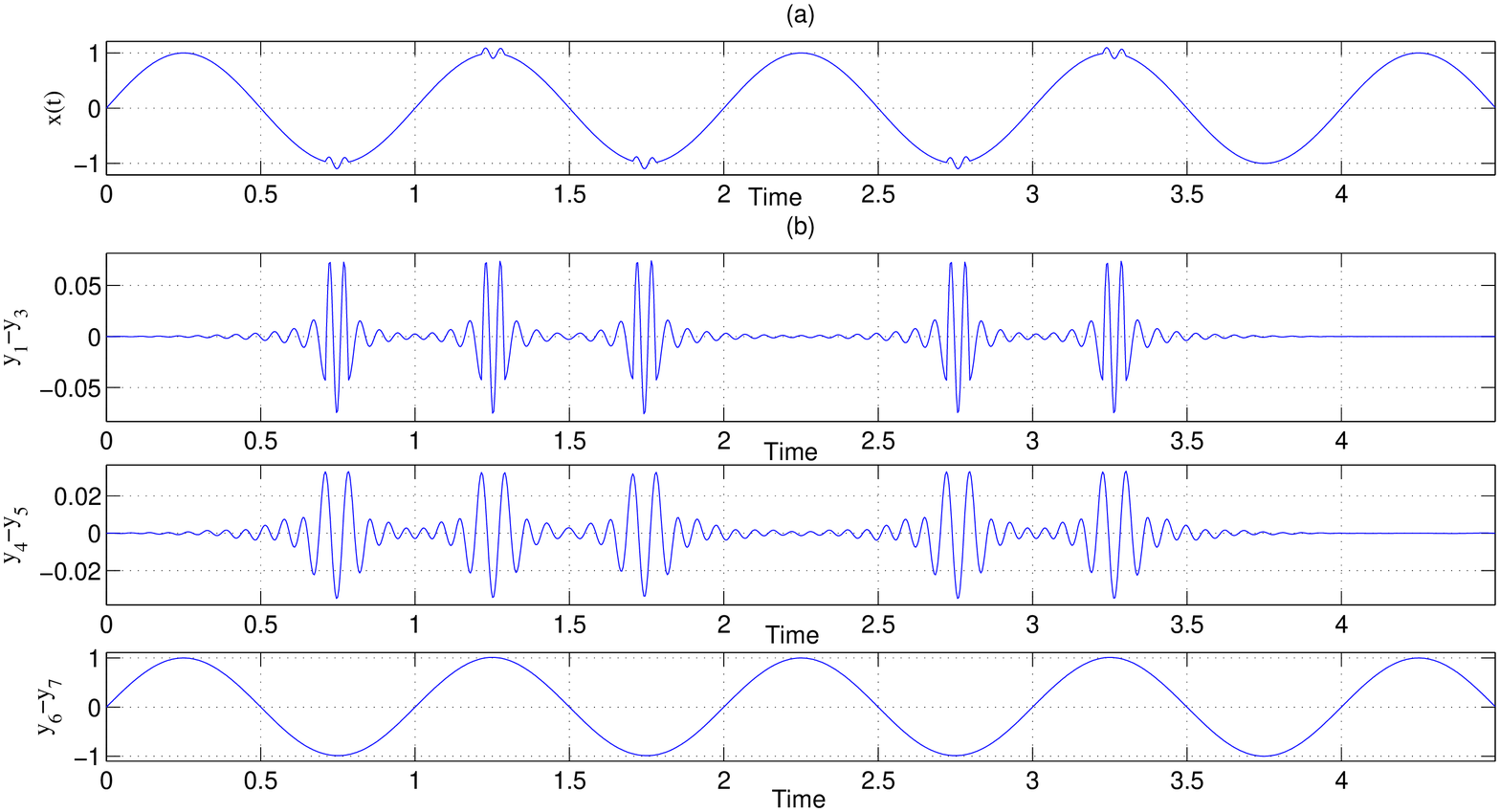}
\includegraphics[angle=0,width=0.5\textwidth,height=0.3\textwidth]{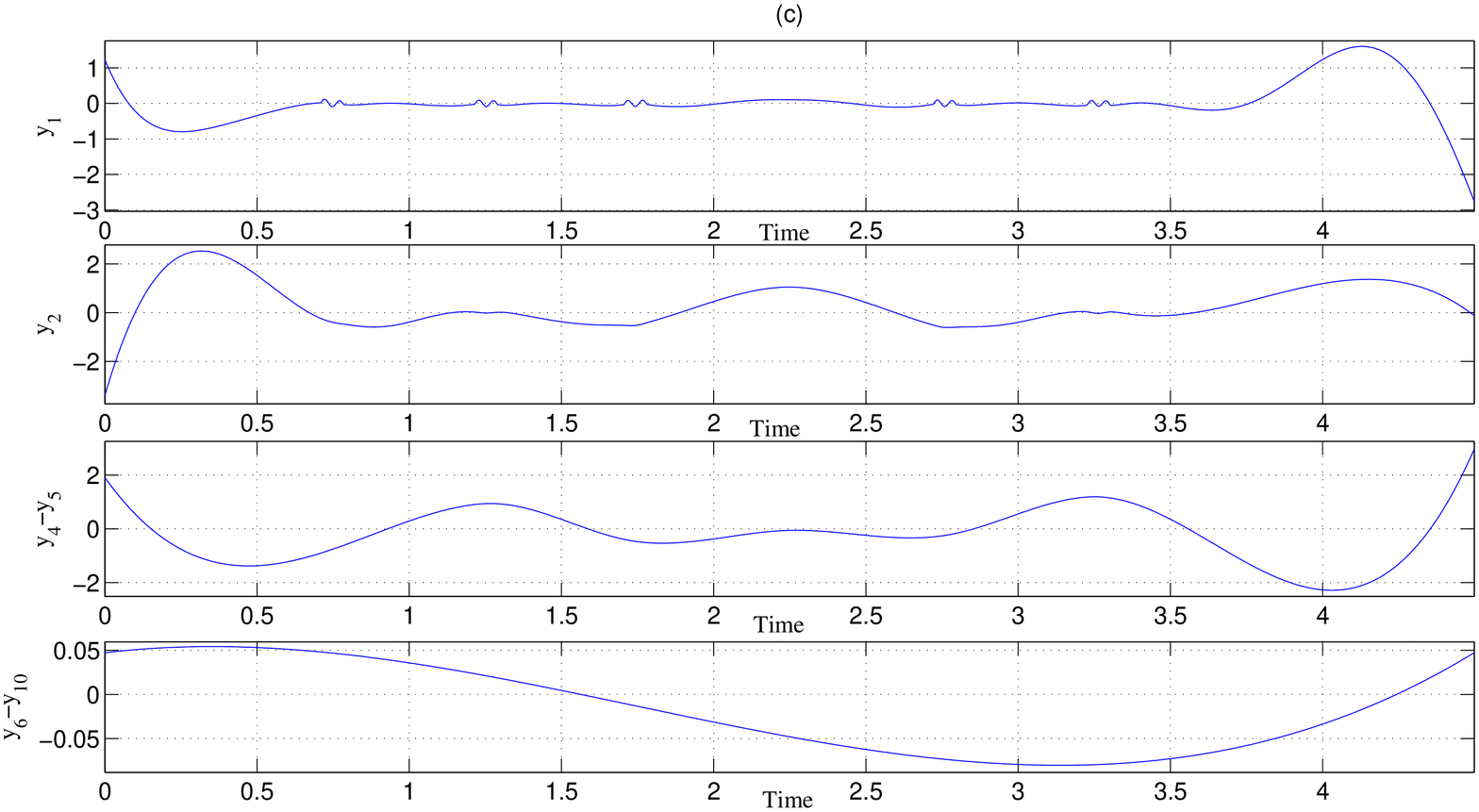}
\includegraphics[angle=0,width=0.5\textwidth,height=0.3\textwidth]{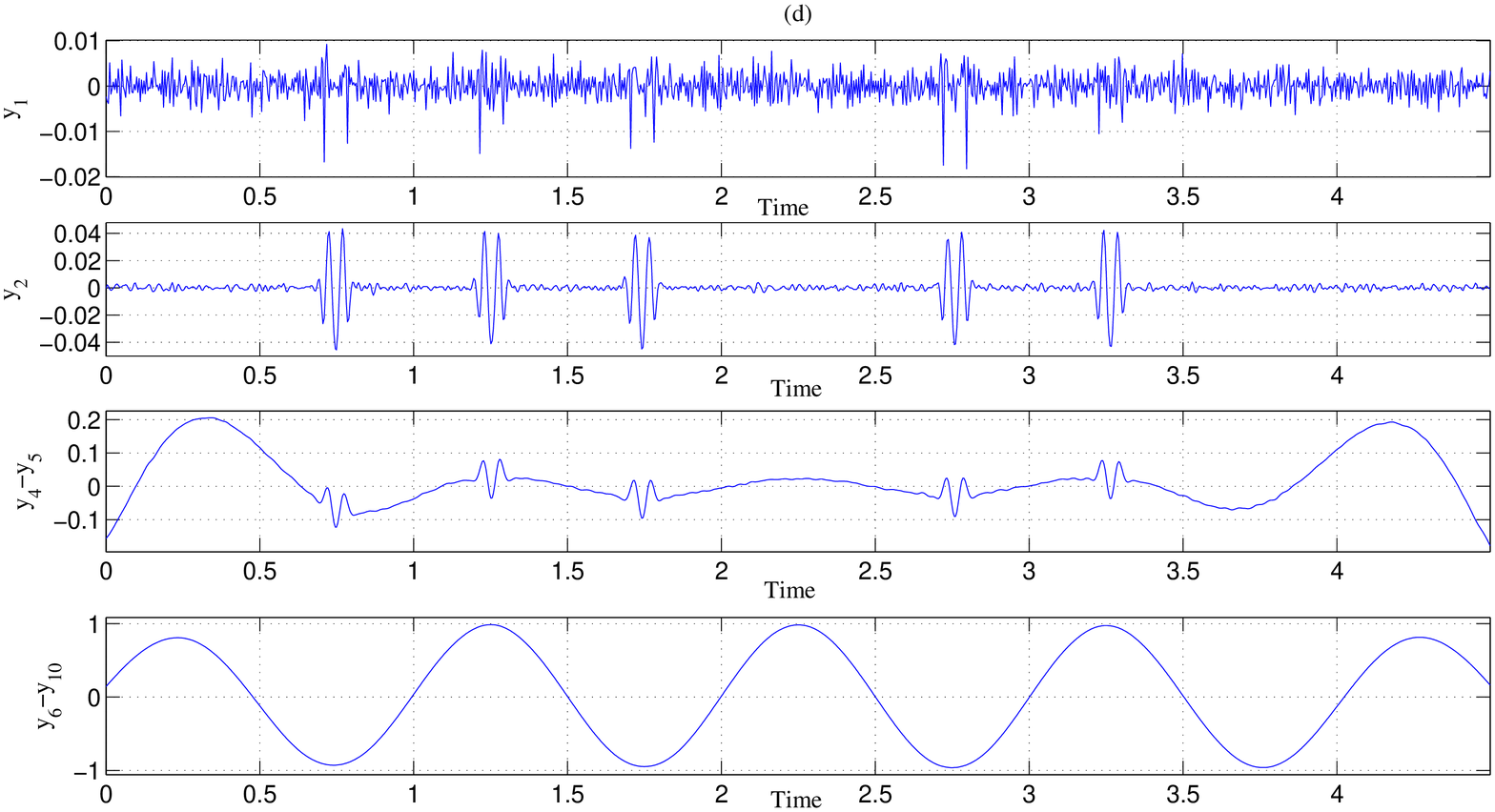}
\caption{Decompositions of a sinusoid mixed with intermittent interference through algorithm of (b) FDM  (c) EMD (d) EEMD.}
\label{fig:modeMix}
\end{figure}
The intermittency, in the time series, is a main cause of mode mixing [e.g. signal $x(t)$ in Figure \ref{fig:modeMix}(a)] and mode splitting in EMD algorithm. These issues are mitigated by EEMD and NA-MEMD. The decompositions of a signal $x(t)$ through FDM algorithm is shown in Figure \ref{fig:modeMix}(b) without end effect artefacts. The MFDM is able to localize the mono-component sinusoid within a single FIBF and outperforming EMD \ref{fig:modeMix}(c) and EEMD \ref{fig:modeMix}(d). On the same machine, computation time for MFDM, EMD and EEMD are 0.56 sec., 0.21 sec. and 77.18 sec., respectively. The ensemble size for EEMD was N = 500 with the 16.94 dB signal-to-noise power ratio.
\subsection{Time-Frequency-Energy Analysis}
\begin{figure}[!t]
\centering
\includegraphics[angle=0,width=0.5\textwidth,height=0.3\textwidth]{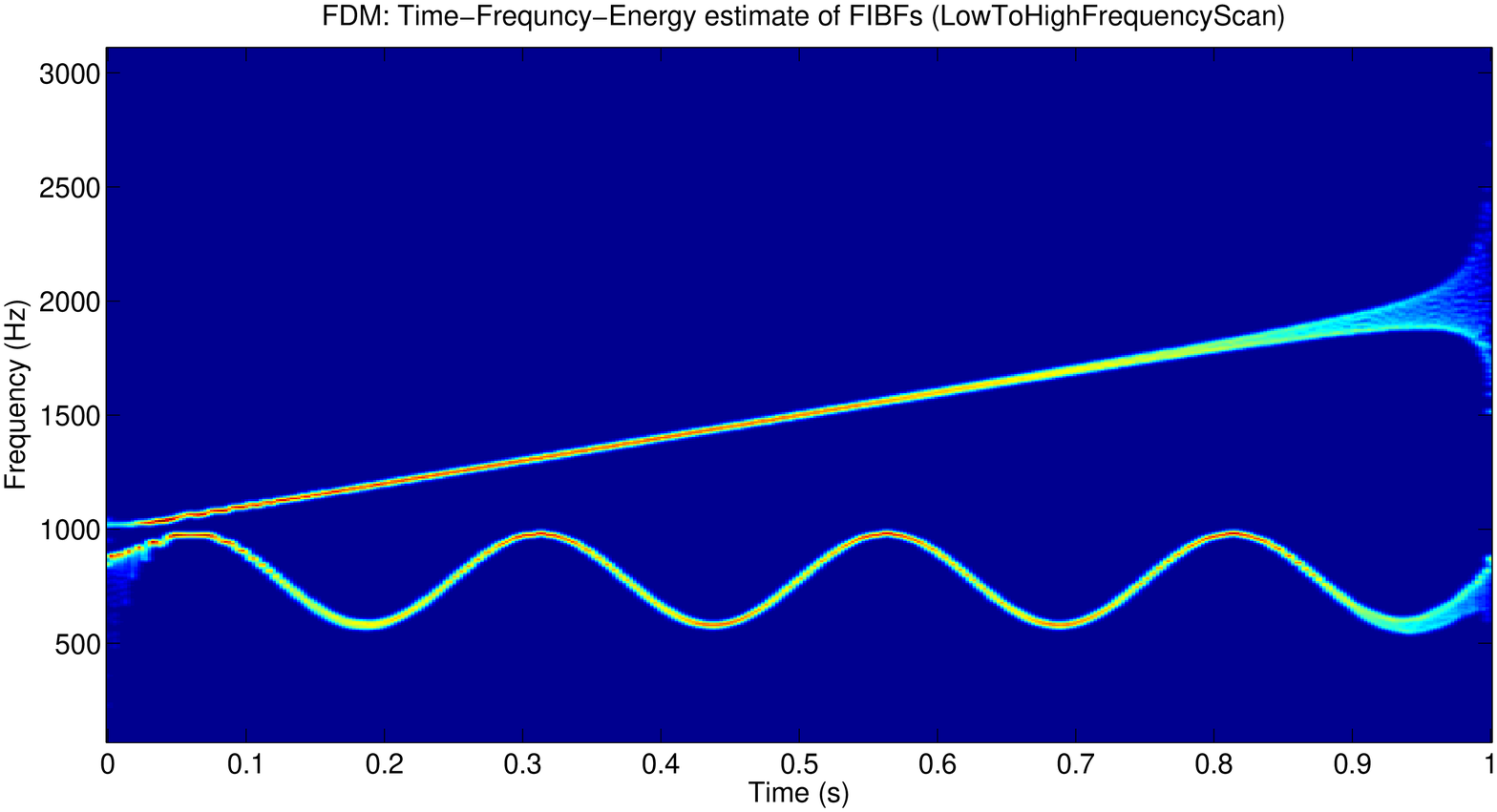}
\includegraphics[angle=0,width=0.5\textwidth,height=0.3\textwidth]{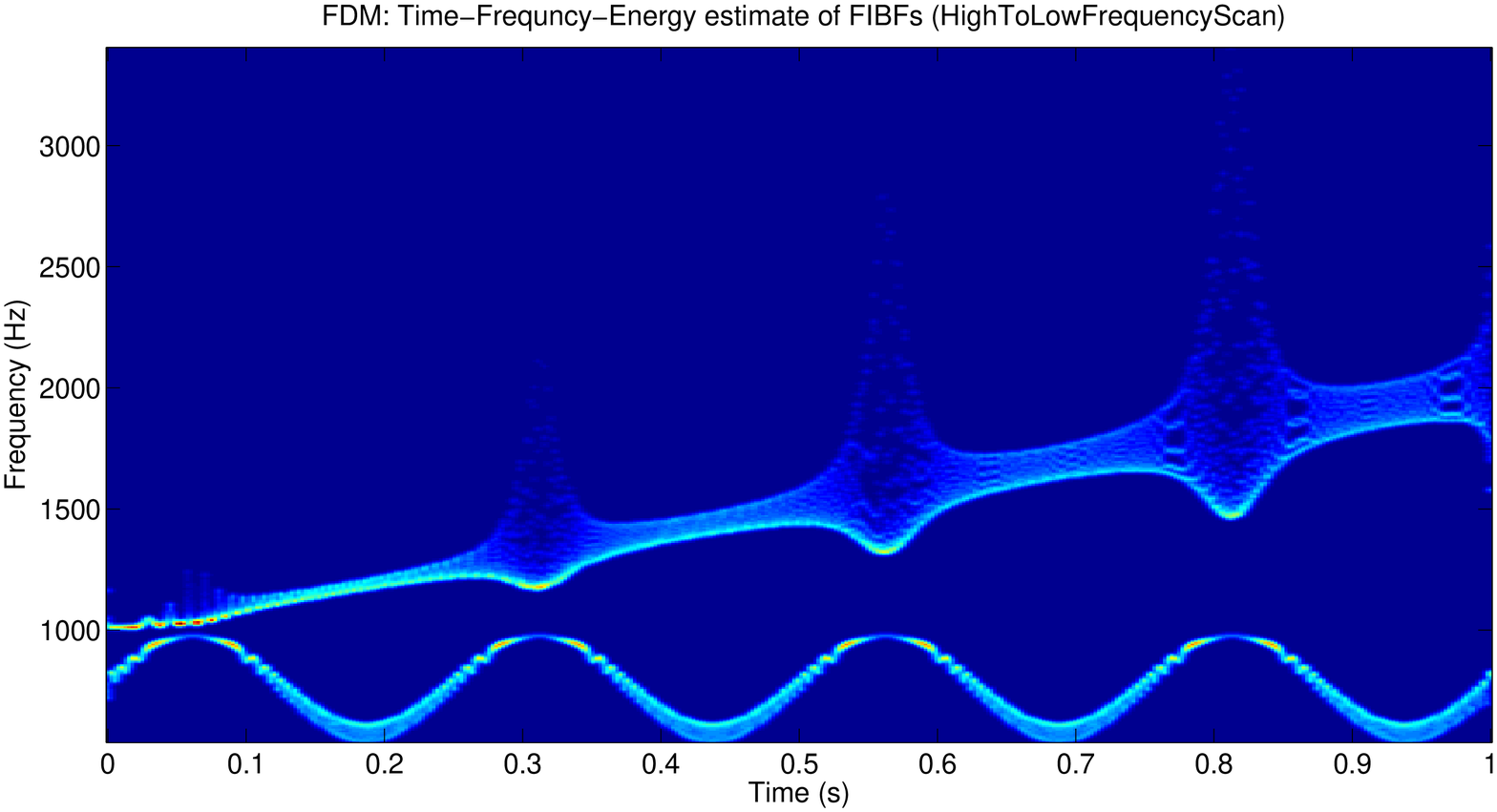}
\includegraphics[angle=0,width=0.5\textwidth,height=0.3\textwidth]{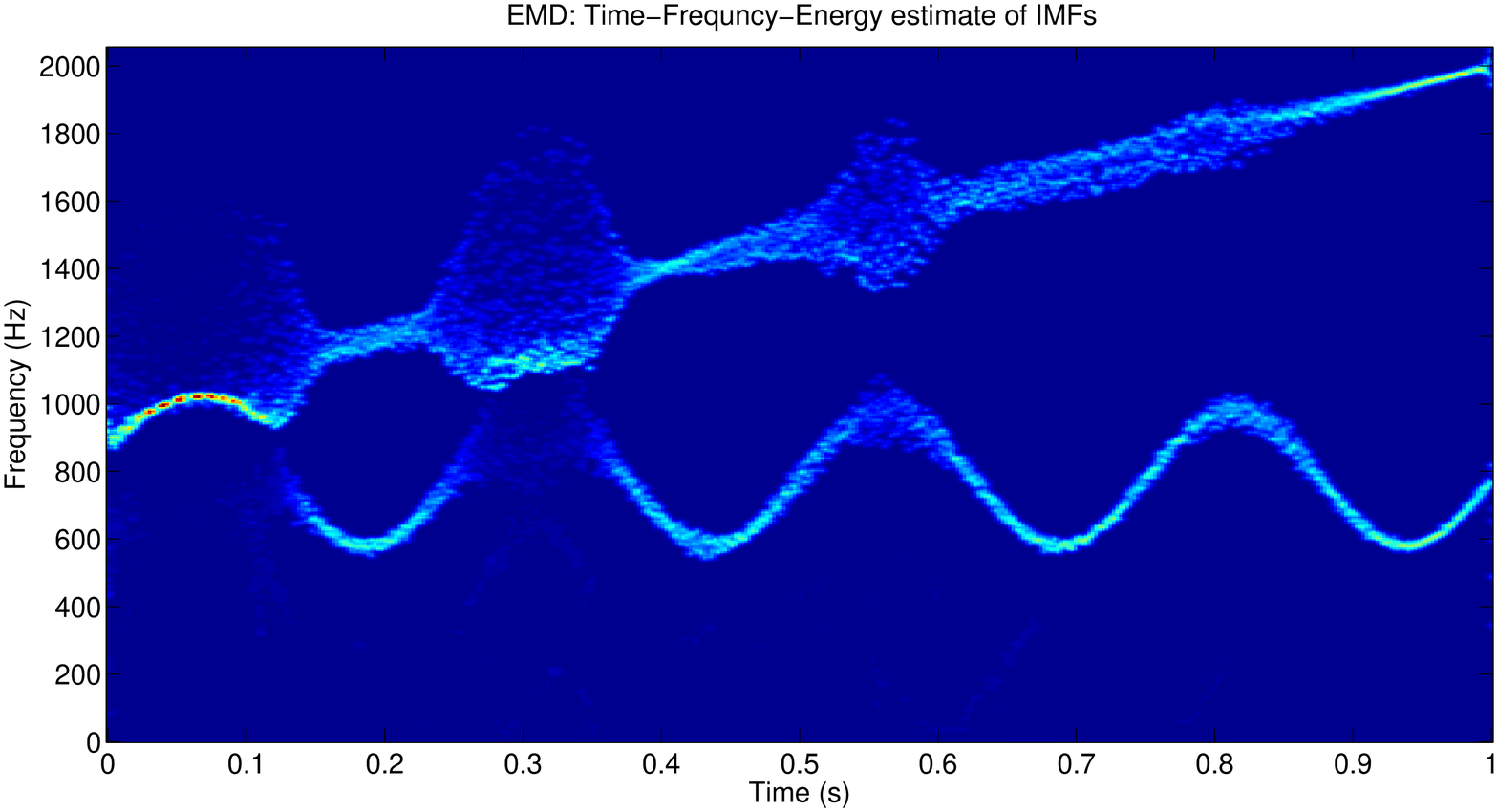}
\captionof{figure}{The TFE analysis of nonstationary signals which is mixture of linear chirp and FM sinusoid: FDM (top and middle) and EMD (bottom).}
\label{fig:TFR_chirpFm}
\end{figure}
Figure~\ref{fig:TFR_chirpFm} shows time-frequency-energy (TFE) estimates for a nonstationary signal mixture of a linear chirp and frequency
modulated (FM) sinusoid, obtained using the FDM and EMD. There is a enhanced TFE tracking when using FDM with low to high frequency scan and other one (HTL-FS) is similar to plot obtained by EMD algorithm.
\subsection{Intrawave frequency modulation}
\begin{figure}[!t]
\centering
\includegraphics[angle=0,width=0.5\textwidth,height=0.3\textwidth]{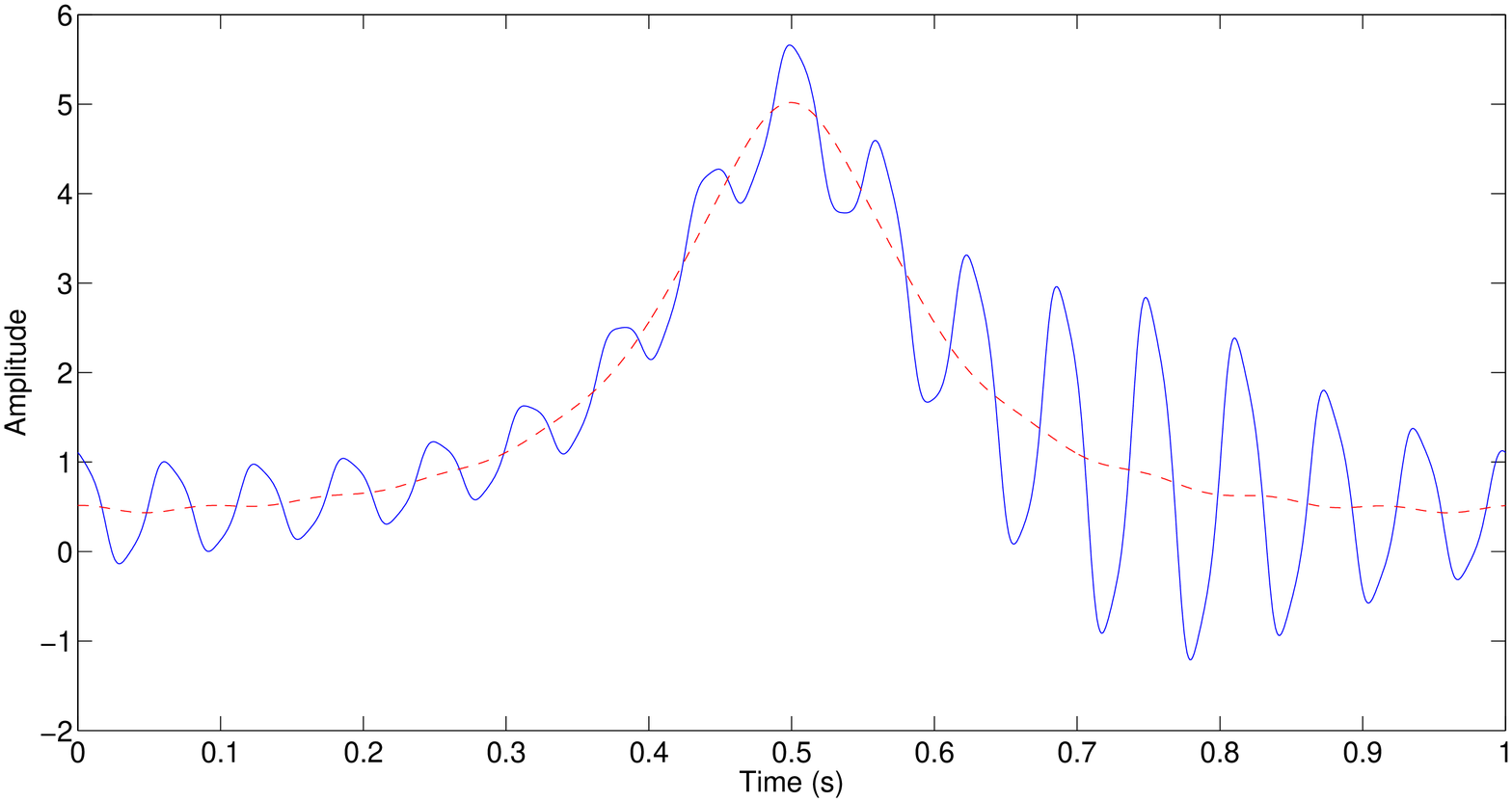}
\captionof{figure}{The signal $x(t)$ (solid line), sum of DC and lowest frequency FIBF (dashed line) of \eqref{IntraWaveMod1}.}
\label{fig:FDM_intraWaveModSig2FIBFs}
\end{figure}
\begin{figure}[!t]
\centering
\includegraphics[angle=0,width=0.5\textwidth,height=0.3\textwidth]{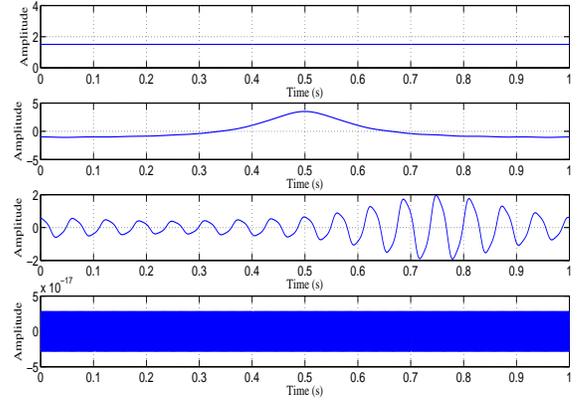}
\captionof{figure}{The DC, FIBF1, FIBF2 and highest frequency component by FDM. The DC and highest frequency component correspond to term $X[0]$ and $X[\frac{N}{2}](-1)^n$ of Eq. \eqref{eq7}, respectively. Sum of DC, FIBF1, FIBF2 and highest frequency component exactly synthesize signal $x(t)$ given by Eq. \eqref{IntraWaveMod1}.}
\label{fig:FDM_intraWaveModSig2FIBFs}
\end{figure}
\begin{figure}[!t]
\centering\
\includegraphics[angle=0,width=0.5\textwidth,height=0.3\textwidth]{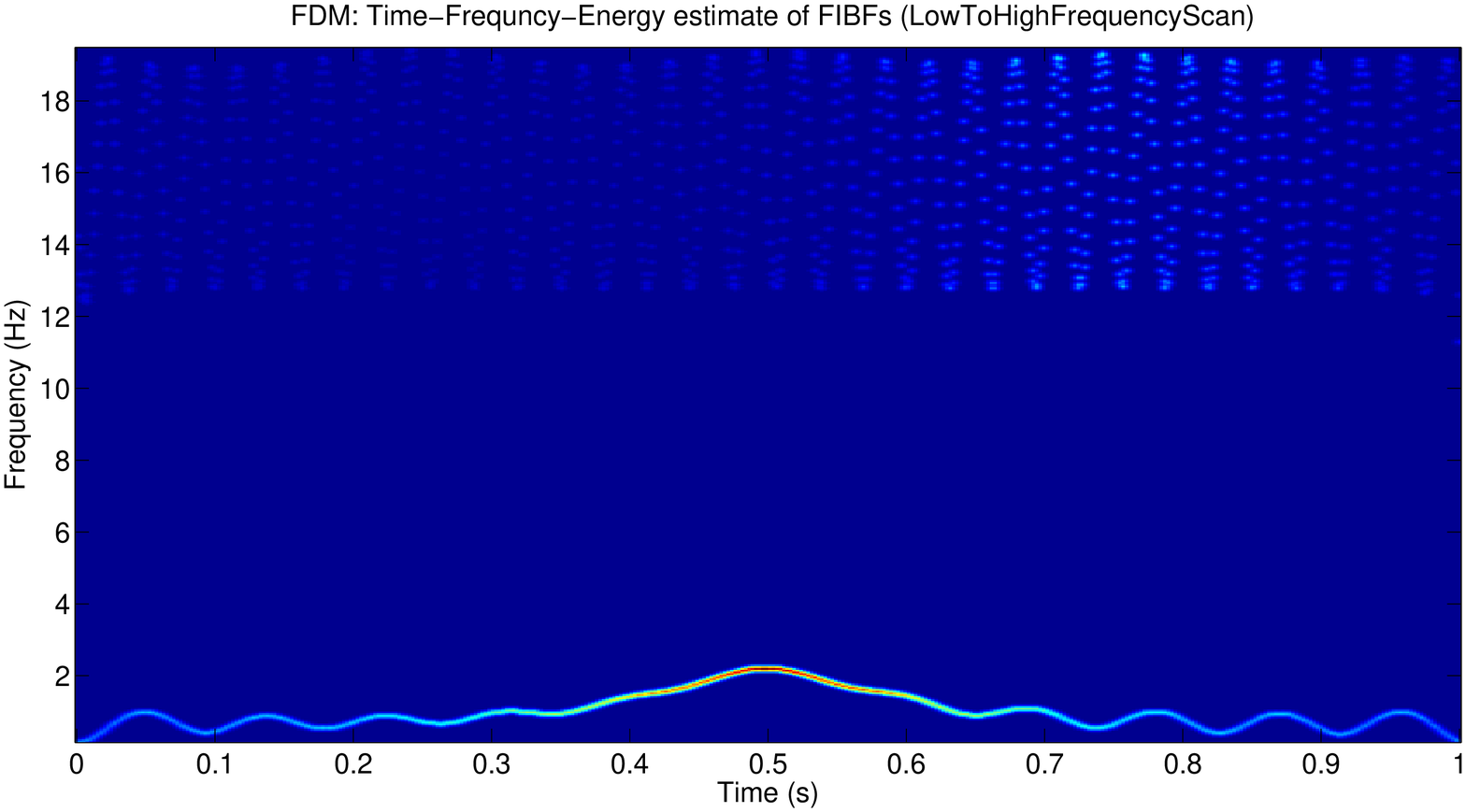}
\includegraphics[angle=0,width=0.5\textwidth,height=0.3\textwidth]{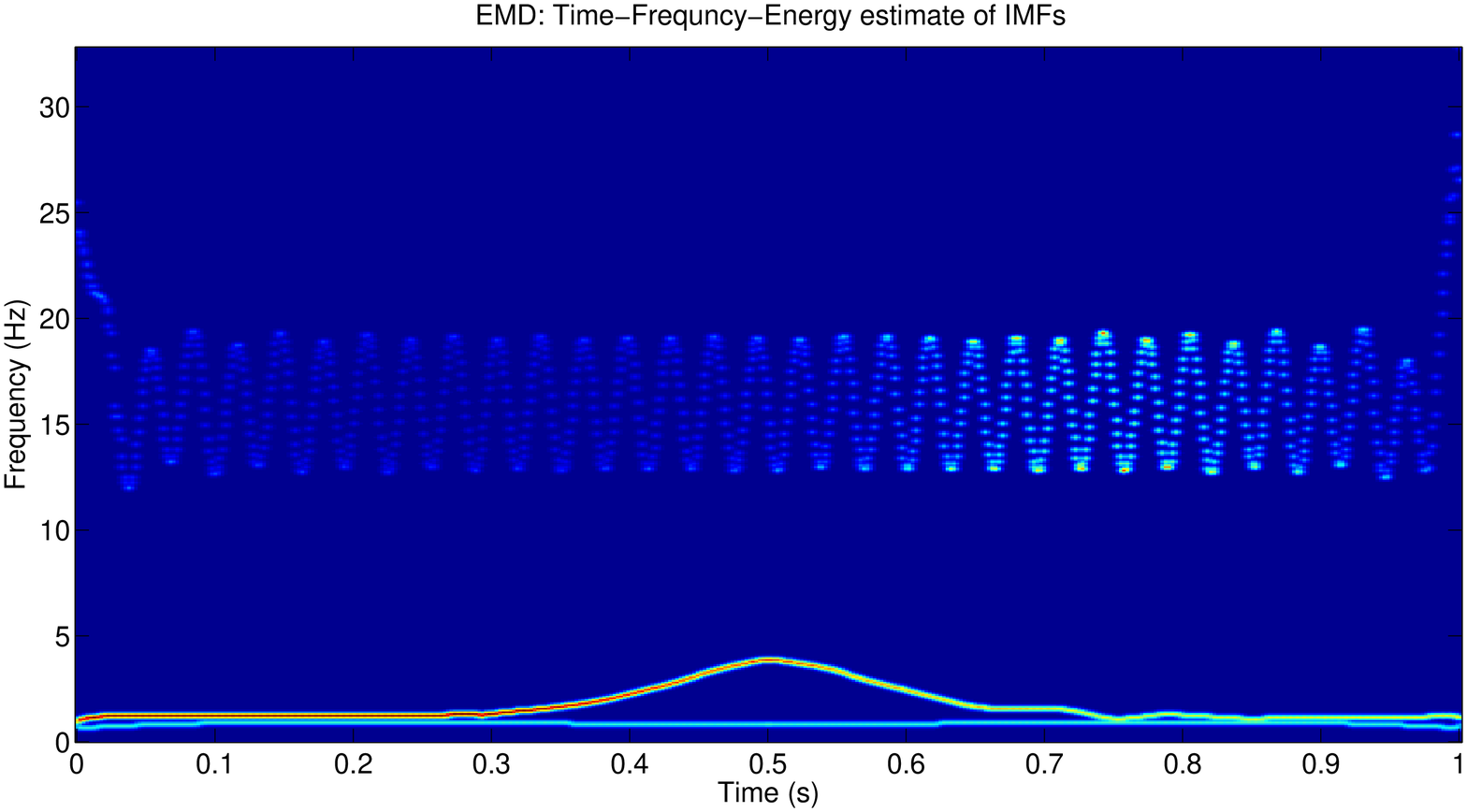}
\captionof{figure}{The TFE analysis of signal $x(t)$ given by Eq. \eqref{IntraWaveMod1}: FDM (top) and EMD (bottom).}
\label{fig:FMD_TFE_intraWaveMod}
\end{figure}
\begin{figure}[!t]
\centering\
\includegraphics[angle=0,width=0.5\textwidth,height=0.3\textwidth]{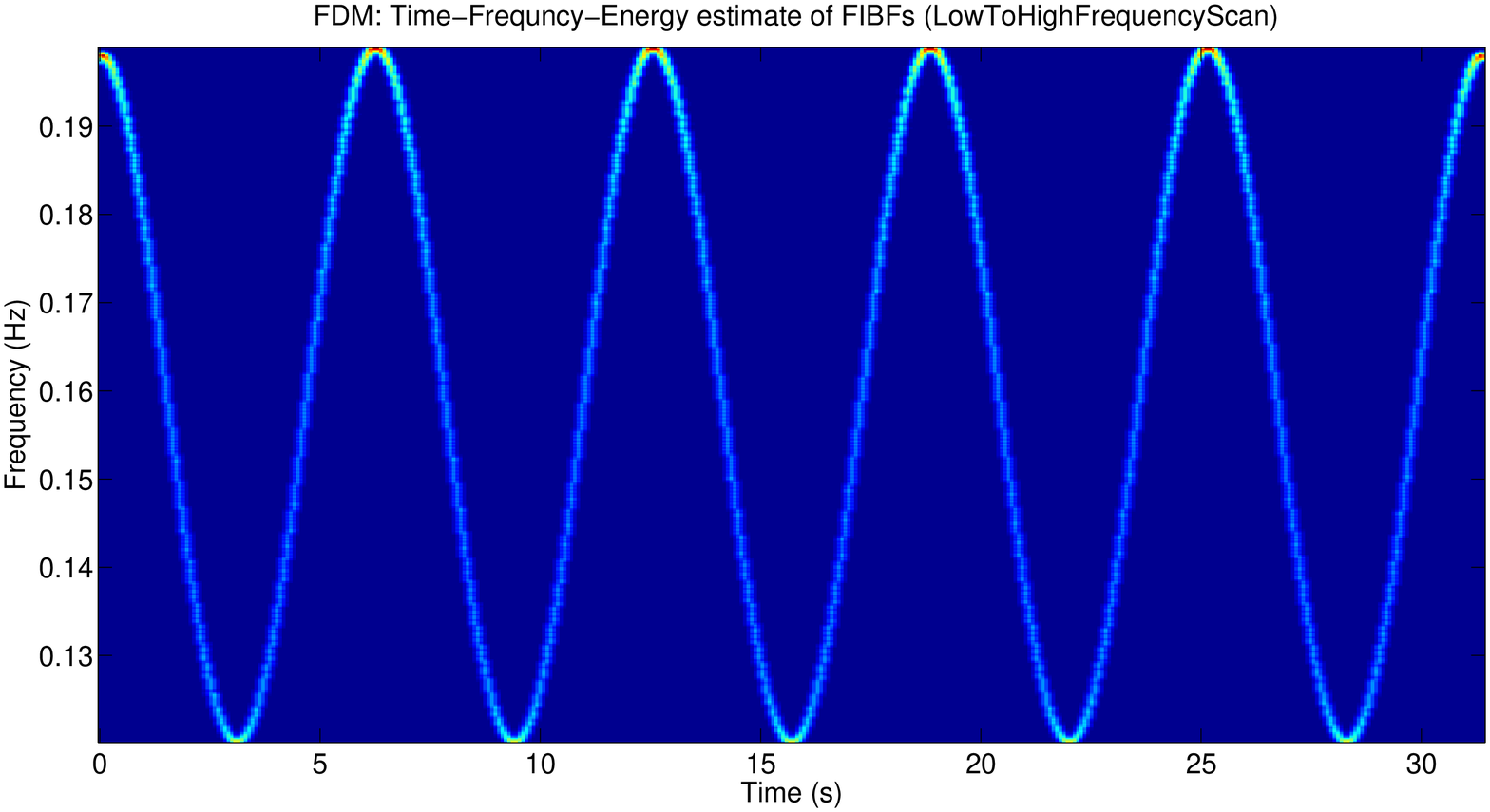}
\includegraphics[angle=0,width=0.5\textwidth,height=0.3\textwidth]{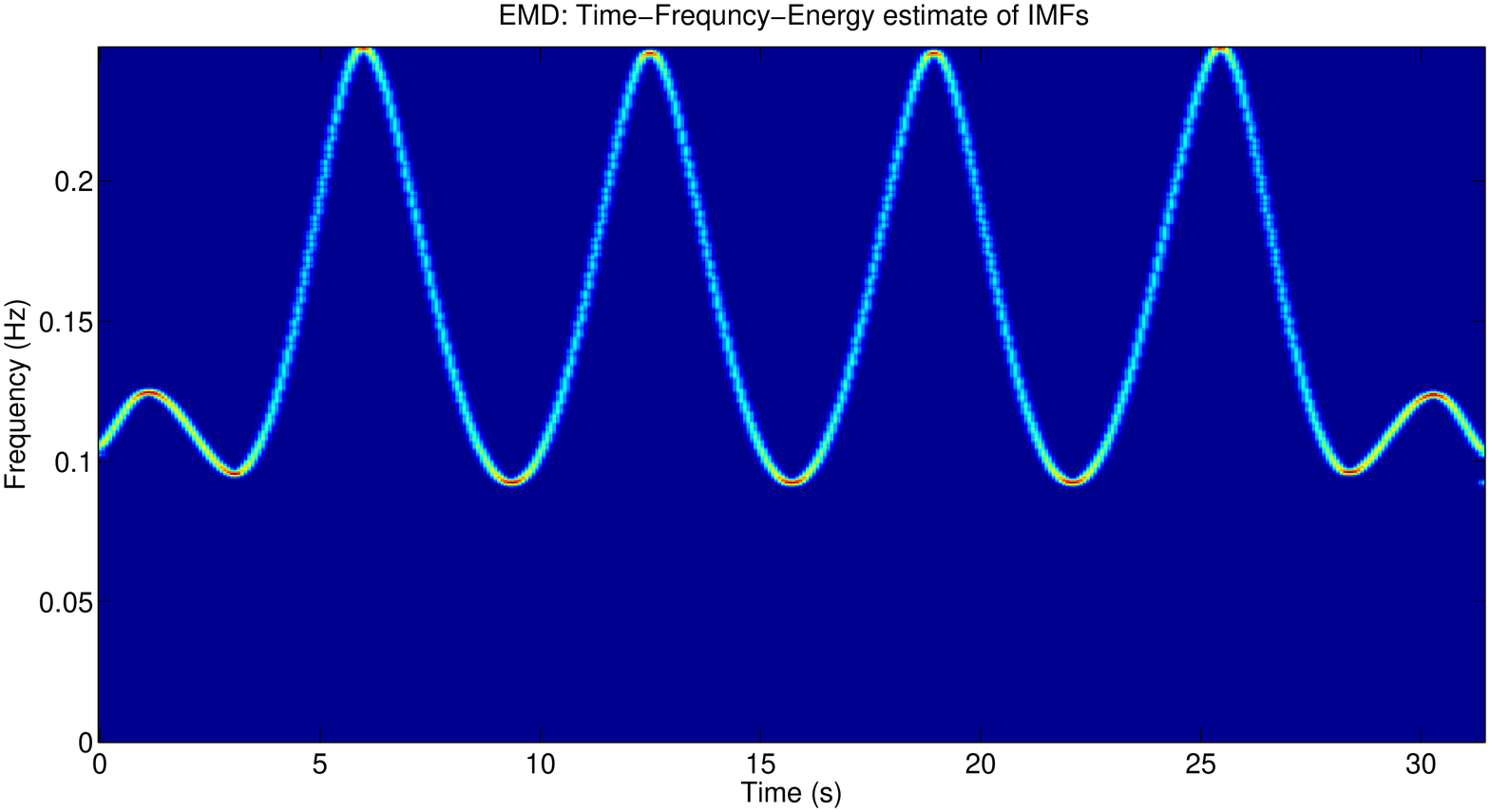}
\captionof{figure}{The TFE analysis of signal $x(t)$ given by Eq. \eqref{IntraWaveMod2}: FDM (top) and EMD (bottom).}
\label{fig:FMD_TFE_intraWaveMod2}
\end{figure}
First, we decompose the following signal that has intrawave frequency modulation and it is considered challenging because the instantaneous frequency itself has very high frequency modulation~\cite{rs32}
\begin{equation}
x(t)=\frac{1}{1.2 + cos(2\pi t)}+\frac{\cos(32 \pi t + 0.2 \cos(64 \pi t))}{1.5 + \sin(2\pi t)} \label{IntraWaveMod1}
\end{equation}
Second, we consider a model wave
\begin{equation}
x(t)=\cos(\omega t + \epsilon \sin \omega t) \label{IntraWaveMod2}
\end{equation} that satisfies the following highly nonlinear differential equation~\cite{rs1}
\begin{equation}
\frac{\ud^2x(t)}{\ud t^2}+[\omega + \epsilon \omega \cos (\omega t)]^2x(t)-[\epsilon \omega^2 \sin( \omega t)]\sqrt{1-x^2(t)}=0 \label{IntraWaveMod21}
\end{equation}
with $\omega=1$ and $\epsilon=0.5$.
We demonstrate that, Figure~\ref{fig:FDM_intraWaveModSig2FIBFs}, \ref{fig:FMD_TFE_intraWaveMod} and~\ref{fig:FMD_TFE_intraWaveMod2}, our method well applies to these challenging cases with good accuracy. These examples clearly demonstrate that the FDM can indeed analyze nonlinear signals and it is a nonlinear decomposition method.
\subsection{Analysis of a white Gaussian noise}
\begin{figure}[!t]
\centering
\includegraphics[angle=0,width=0.5\textwidth,height=0.3\textwidth]{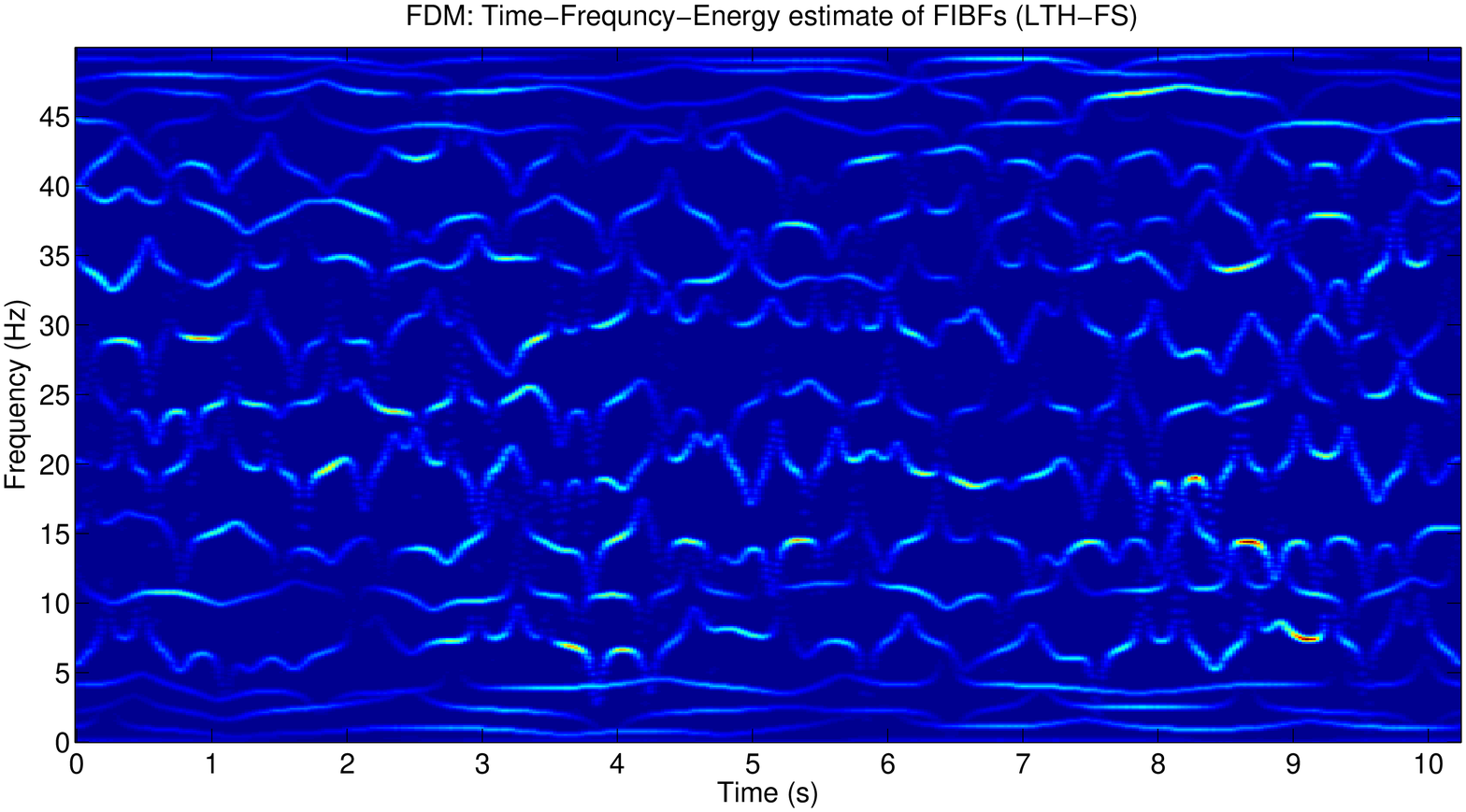}
\includegraphics[angle=0,width=0.5\textwidth,height=0.3\textwidth]{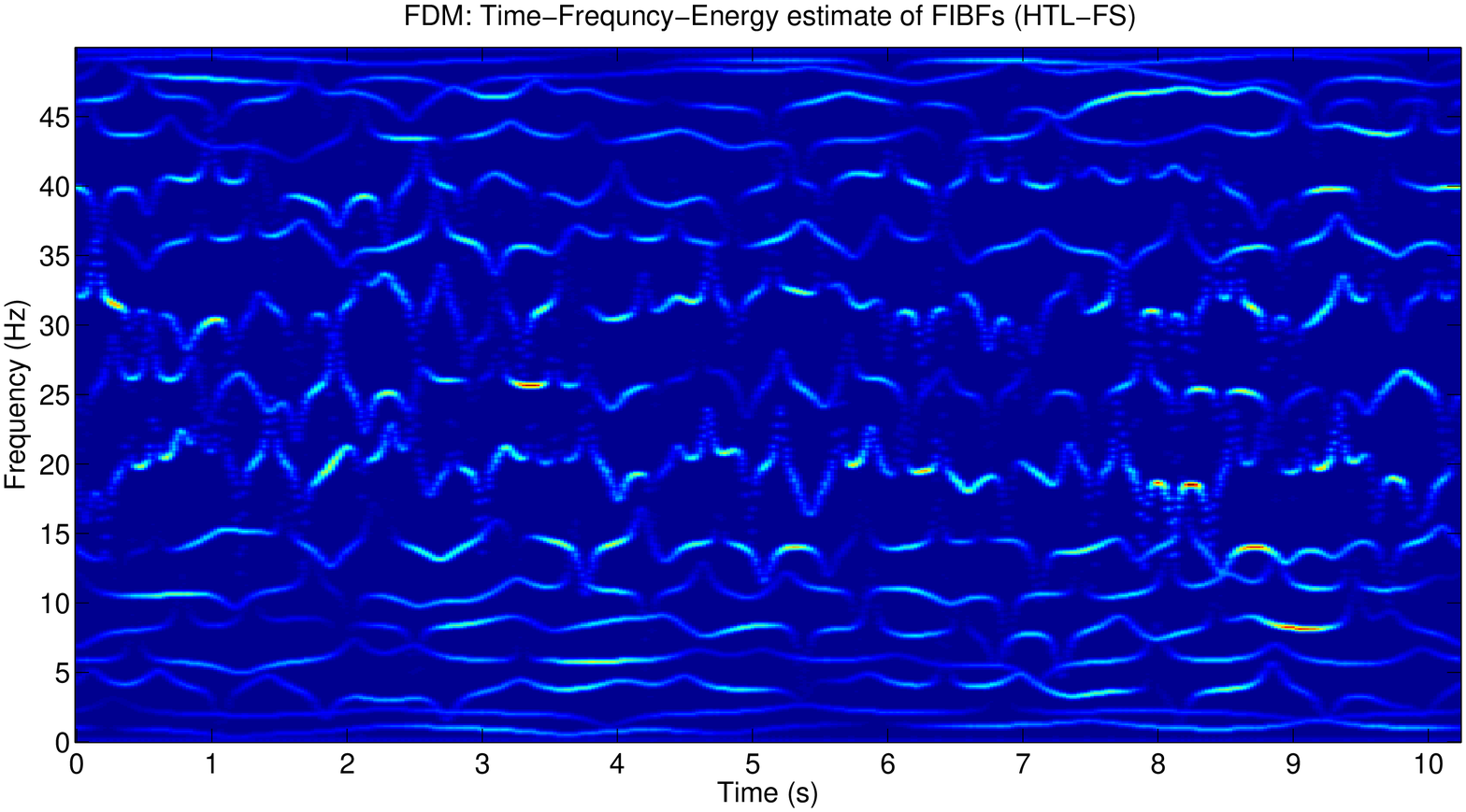}
\includegraphics[angle=0,width=0.5\textwidth,height=0.3\textwidth]{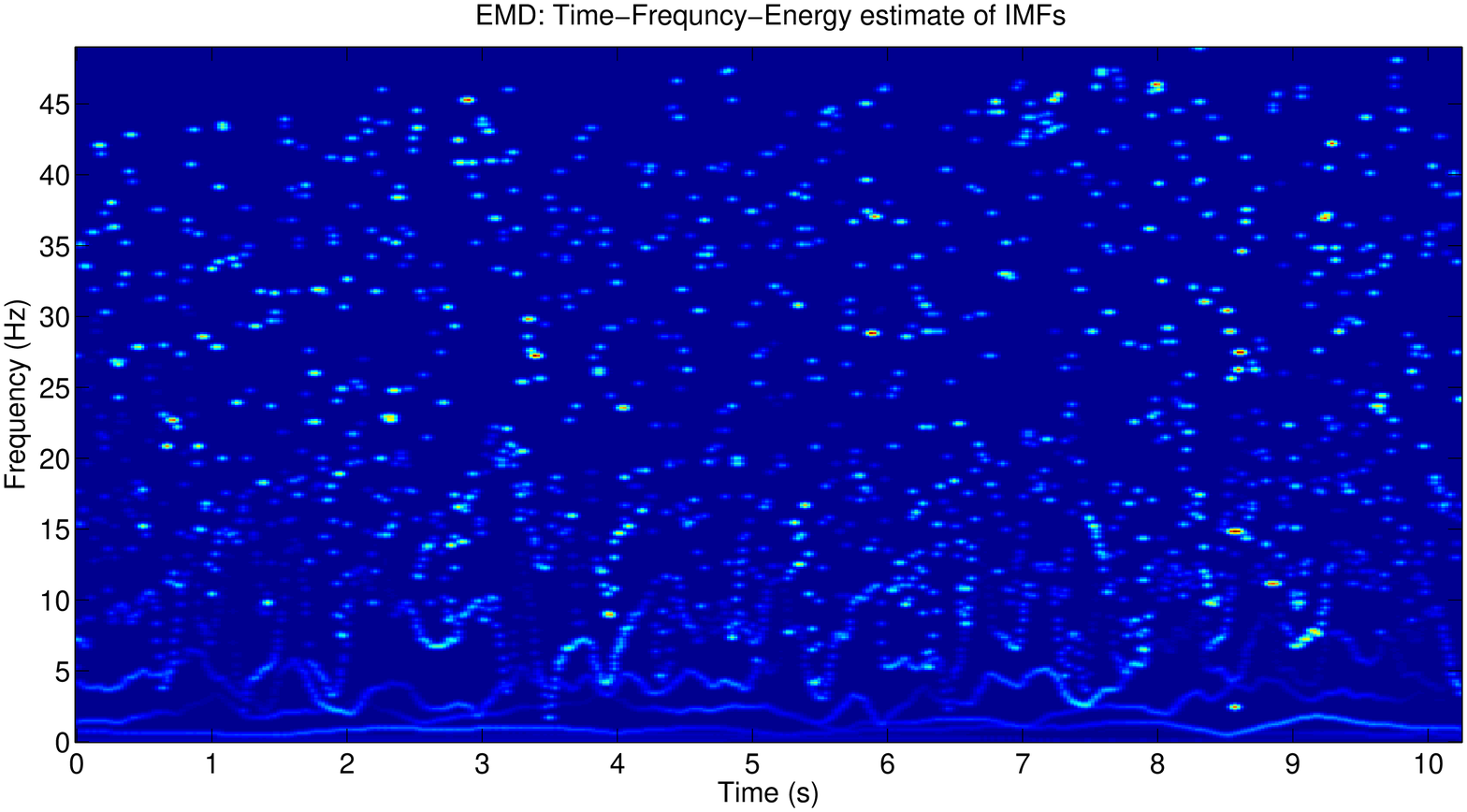}
\captionof{figure}{The TFE analysis of the Gaussian white noise (with zero mean and unit variance): FDM (top and middle) and EMD (bottom).}
\label{fig:WGN_TFE}
\end{figure}
\begin{figure}[!t]
\centering
\includegraphics[angle=0,width=0.5\textwidth,height=0.3\textwidth]{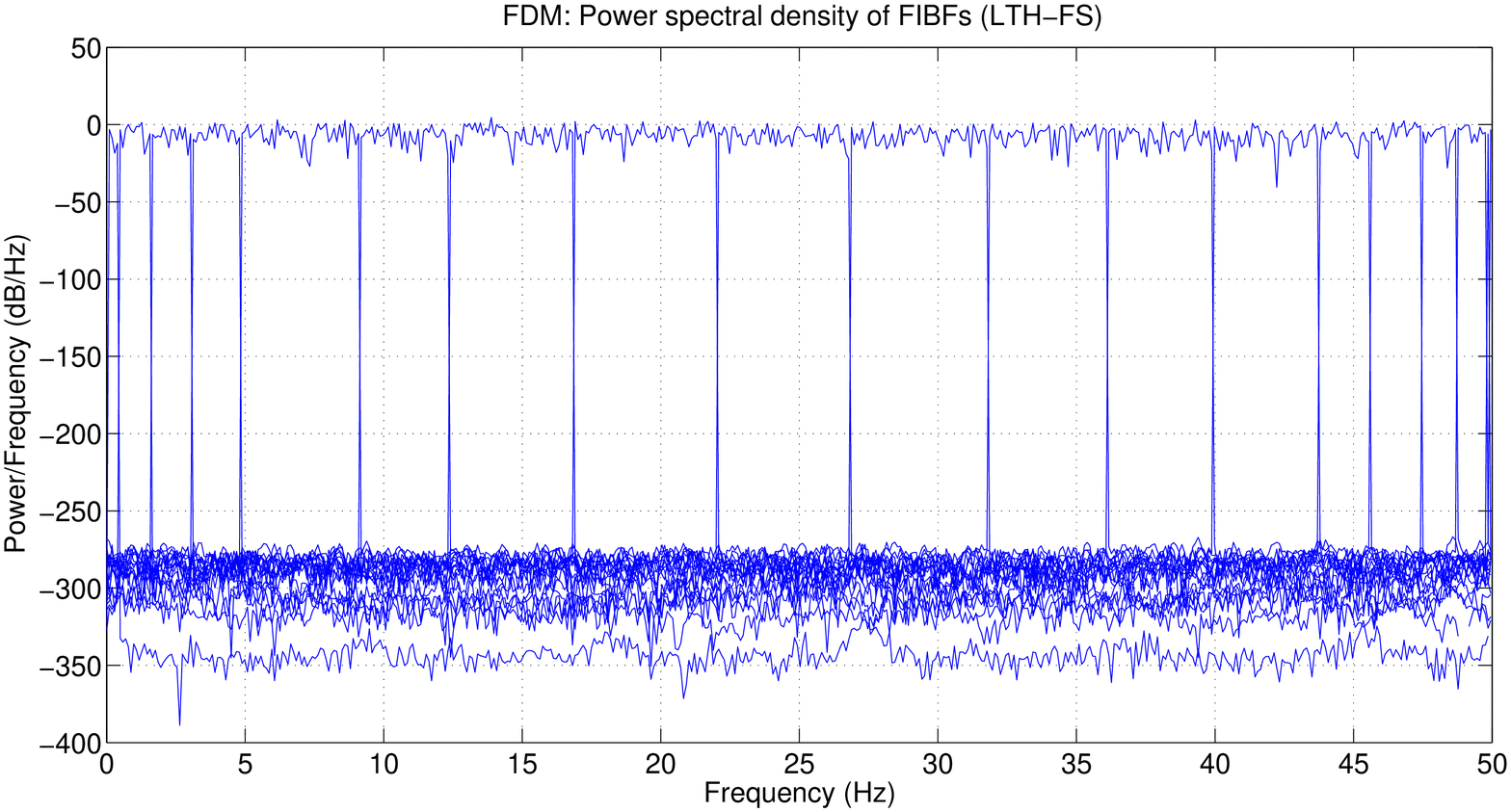}
\includegraphics[angle=0,width=0.5\textwidth,height=0.3\textwidth]{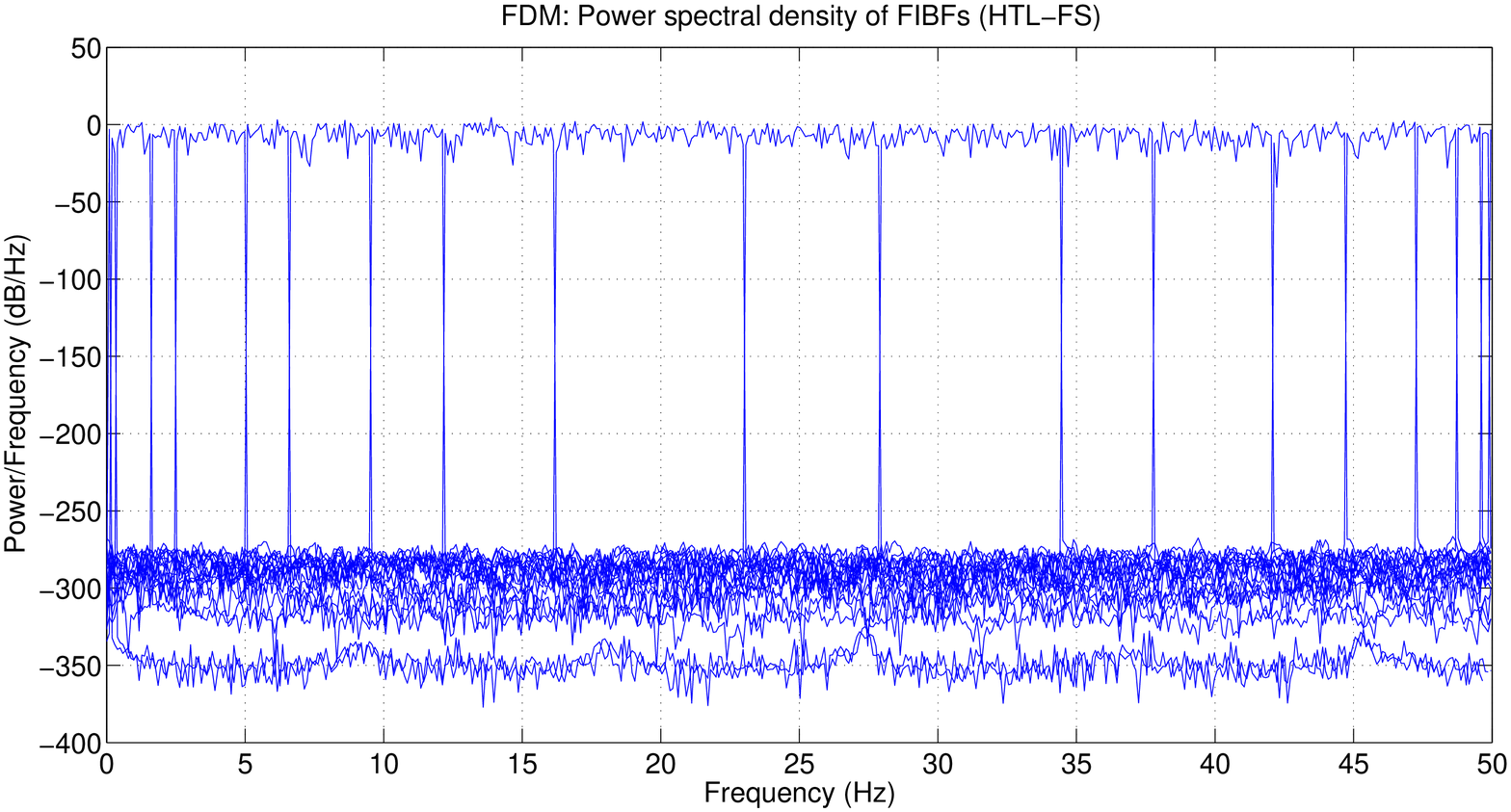}
\includegraphics[angle=0,width=0.5\textwidth,height=0.3\textwidth]{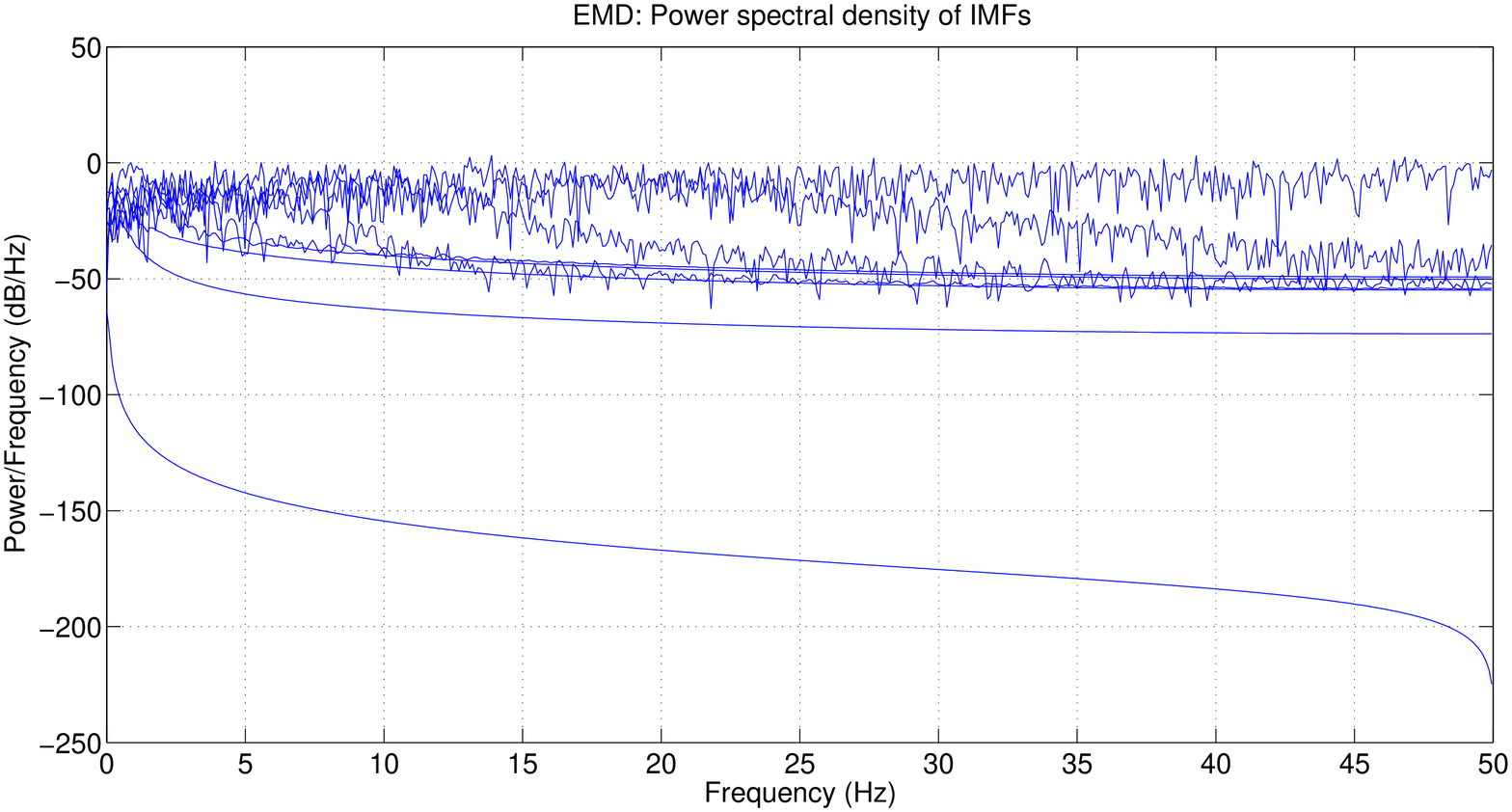}
\captionof{figure}{The PSD analysis of the Gaussian white noise (with zero mean and unit variance): FDM (top and middle) and EMD (bottom).}
\label{fig:WGN_PSD}
\end{figure}
Figure~\ref{fig:WGN_TFE} shows the TFE analysis of a white Gaussian noise (with zero mean, unit variance, 1024 samples and sampling frequency $F_s=100$ Hz) obtained from the FDM and EMD algorithm. Clearly, both LTH-FS and HTL-FS views of TFE is similar and complete data is decomposed in FIBFs. Figure~\ref{fig:WGN_PSD} shows the power spectral density (PSD) plot of same white Gaussian noise with the FDM and EMD algorithm. The FDM has dived the complete data in narrowband and orthogonal FIBFs. Both LTH-FS and HTL-FS views of PSD looking similar but FIBFs have different frequency band, e.g. approximately 40 Hz is cutoff frequency for one of the band in PSD (LTH-FS), whereas, it is mid frequency of the one band in other PSD (HTL-FS) view. There are enhanced TFE and PSD tracking when using FDM.
\subsection{TFE Analysis of unit sample sequence}
\begin{figure}[!t]
\centering
\includegraphics[angle=0,width=0.5\textwidth,height=0.3\textwidth]{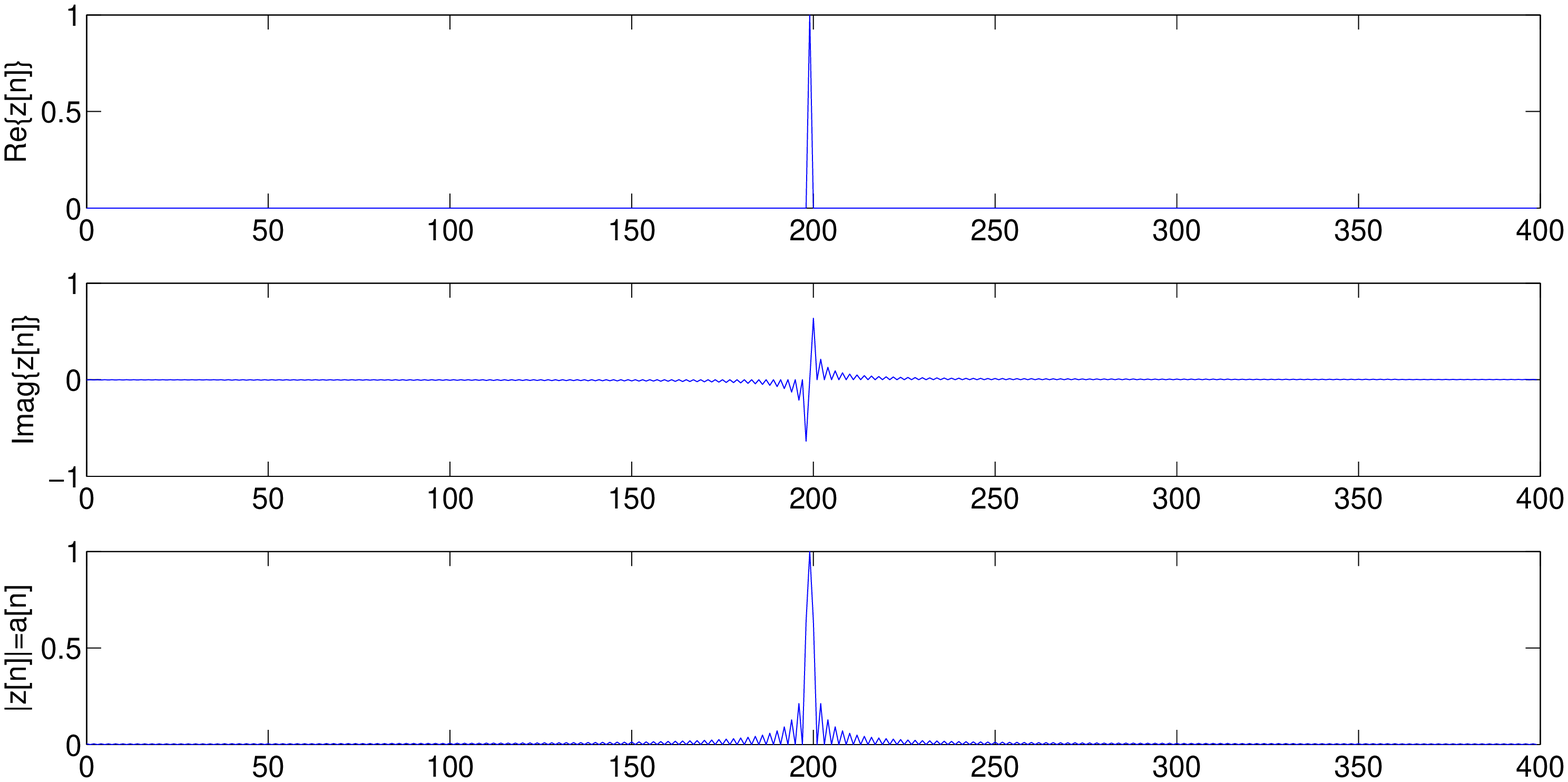}
\captionof{figure}{The analytic representation of $\delta [n-n_0]$ (with, $n_0=199$, sampling frequency $F_s=100$ Hz, length $N=400$): Real part of $z[n]$ (top), Imaginary part of $z[n]$ (middle) and absolute value of $z[n]$ (bottom).}
\label{fig:uss_zn}
\end{figure}
The unit sample sequence is defined as $\delta [n-n_0]=1$ at $n=n_0$ and zero otherwise. By using the relation $z[n]=\frac{1}{\pi}\int_{0}^{\pi} X(\omega) \exp(j\omega n) \ud \omega$, we obtain the analytic representation of $x[n]=\delta [n-n_0]\Leftrightarrow X(\omega)=\exp(-j\omega n_0)$ as
\begin{equation}
 \begin{aligned}
        z[n] & =\frac{\sin(\pi (n-n_0))+j[1-\cos(\pi (n-n_0))]}{\pi (n-n_0)}\\
        z[n] & =a[n]\exp(j\phi [n]),
       \end{aligned}
 \label{uss_eq1}
\end{equation}
where real part of $z[n]$ is $\delta [n-n_0]=\frac{\sin(\pi (n-n_0))}{\pi (n-n_0)}$, $a[n]=\lvert\frac{ \sin(\frac{\pi}{2} (n-n_0))} {\frac{\pi}{2} (n-n_0))}\rvert$, $\phi[n]=\frac{\pi}{2} (n-n_0)$ and hence $\omega[n]=\frac{\pi}{2}$ which corresponds to half of the Nyquist frequency, i.e. $\frac{F_s}{4}$ Hz. Figure~\ref{fig:uss_zn} shows the plots of real, imaginary part and absolute value of $z[n]$ with $n_0=199$, sampling frequency $F_s=100$ Hz, length $N=400$. Theoretically, this clearly indicate that most of the energy of signal $\delta [n-n_0]$ is concentrated at time $t=1.99$ sec. ($n_0=199$) and frequency $f=25$ Hz.
\begin{figure}[!t]
\centering
\includegraphics[angle=0,width=0.5\textwidth,height=0.3\textwidth]{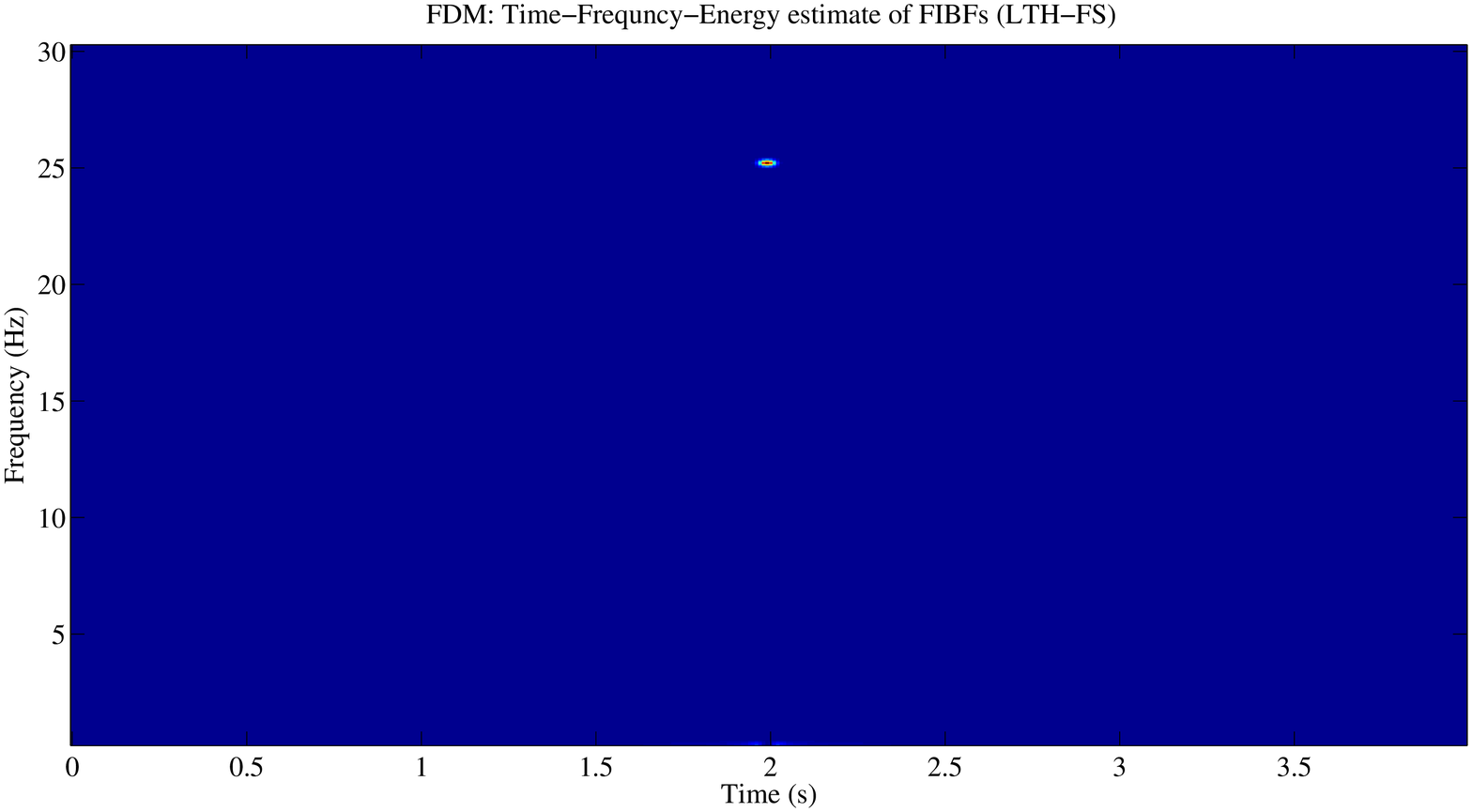}
\includegraphics[angle=0,width=0.5\textwidth,height=0.3\textwidth]{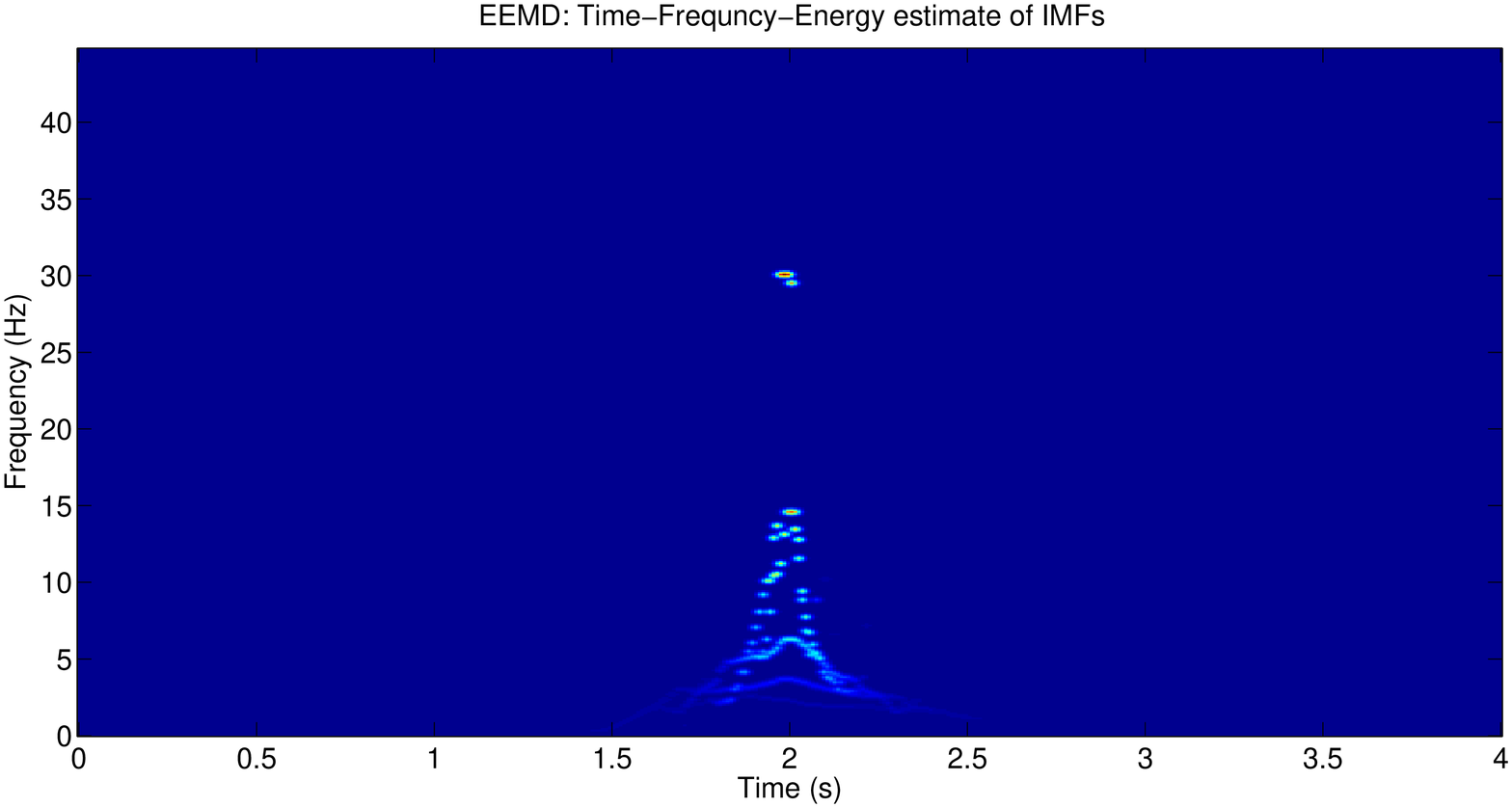}
\includegraphics[angle=0,width=0.5\textwidth,height=0.3\textwidth]{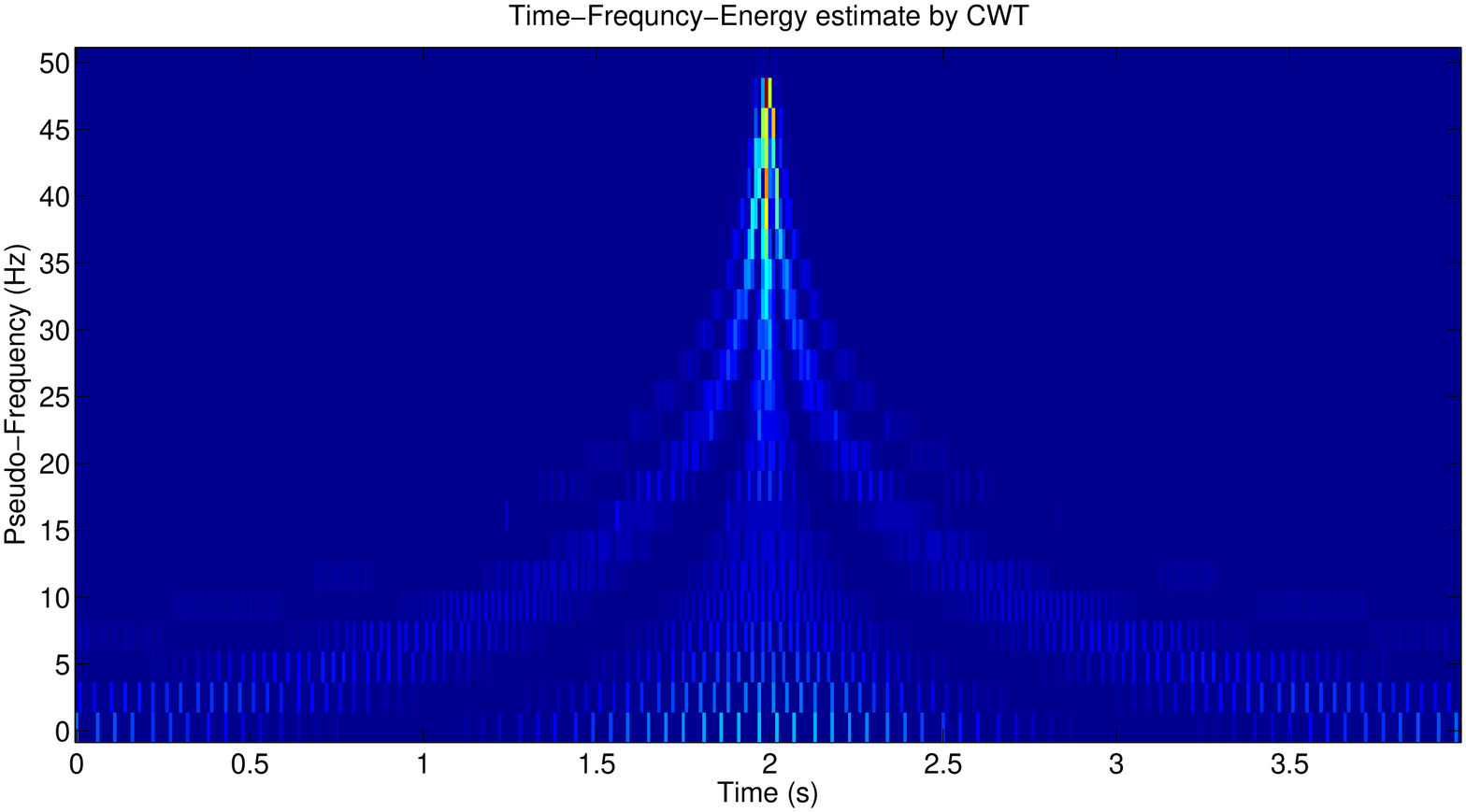}
\captionof{figure}{The TFE analysis of unit sample sequence $\delta [n-n_0]$ (with, $n_0=199$, sampling frequency $F_s=100$ Hz, length $N=400$): FDM (top), EEMD (middle) and CWT (bottom).}
\label{fig:uss_TFE}
\end{figure}
\begin{figure}[!t]
\centering
\includegraphics[angle=0,width=0.5\textwidth,height=0.3\textwidth]{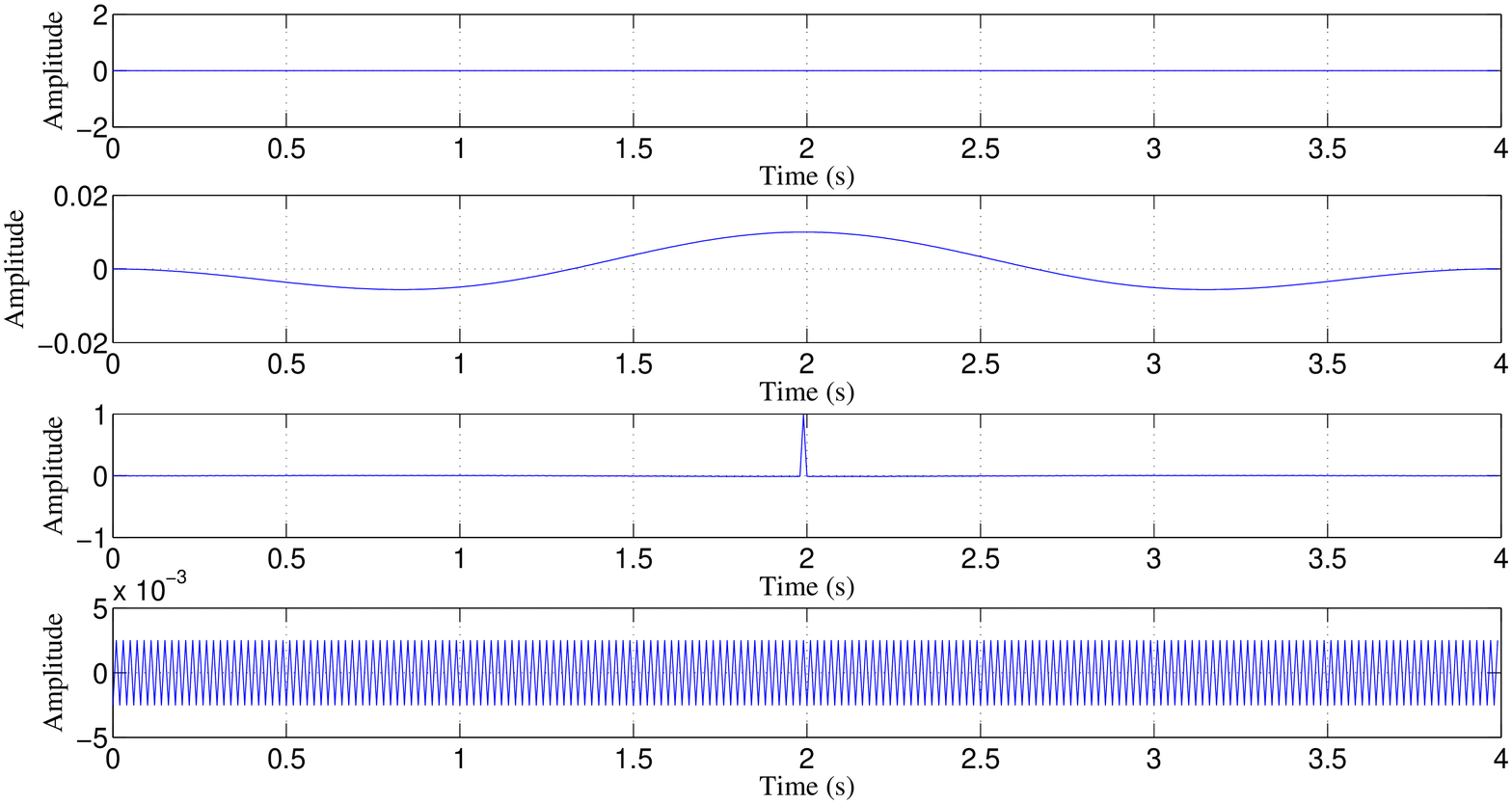}
\captionof{figure}{The DC, FIBF1, FIBF2 and highest frequency component plots by the FDM of unit sample sequence $\delta [n-n_0]$ (with, $n_0=199$, sampling frequency $F_s=100$ Hz, length $N=400$).}
\label{fig:uss_FIBFs}
\end{figure}
 Figure~\ref{fig:uss_FIBFs} shows the plots of FIBFs obtained by the FDM and Figure~\ref{fig:uss_TFE} shows the TFE analysis of unit sample sequence $\delta [n-n_0]$ from the FDM, EEMD and continuous wavelet transform (CWT) methods. This clearly indicate that the TFE plot obtained by the FDM method is same as theoretical estimation, whereas there is energy spread over a range of frequencies and lack of accuracy in TFE plot by the EEMD and CWT methods.

 The uncertainty principle is a consequence of the Fourier transform (or any other type of integral transform) pair in the frequency definition. Therefore, its limitation could only be applied to such integral transforms, in which time would be `integrated out' or smeared over the integral time limit. Consequently, if we avoid an integral transform in the frequency computation, time-frequency analysis of signal would not be bounded by the uncertainty principle~\cite{ucrtp}. The IF is defined through differentiation of phase rather than integration and, hence, overcome the restriction of the uncertainty principle. One can obtain arbitrary precision on time and frequency resolution subject only to the sampling rate in case of discrete time signal.
 The uncertainty principle in signal analysis state that the finer the time resolution one wants, the cruder the resulting frequency resolution would be. However, this example clearly demonstrate that the FHS by FDM is indeed not limited by uncertainty principle and signal can be highly concentrated in both time and frequency domain.
 \subsection{Application to a earthquake signal analysis}
\begin{figure}[!t]
\centering
\includegraphics[angle=0,width=0.5\textwidth,height=0.3\textwidth]{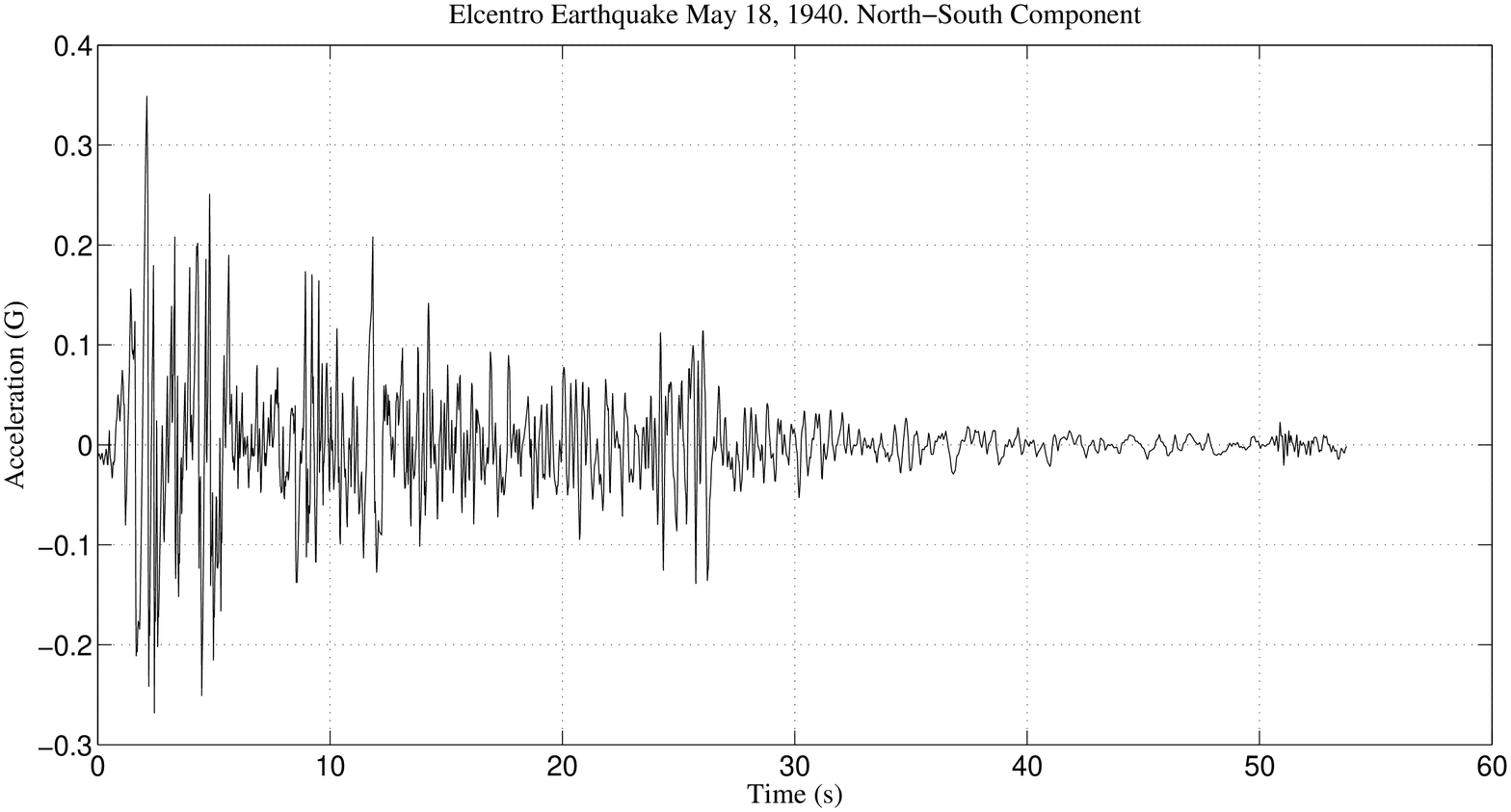}
\includegraphics[angle=0,width=0.5\textwidth,height=0.3\textwidth]{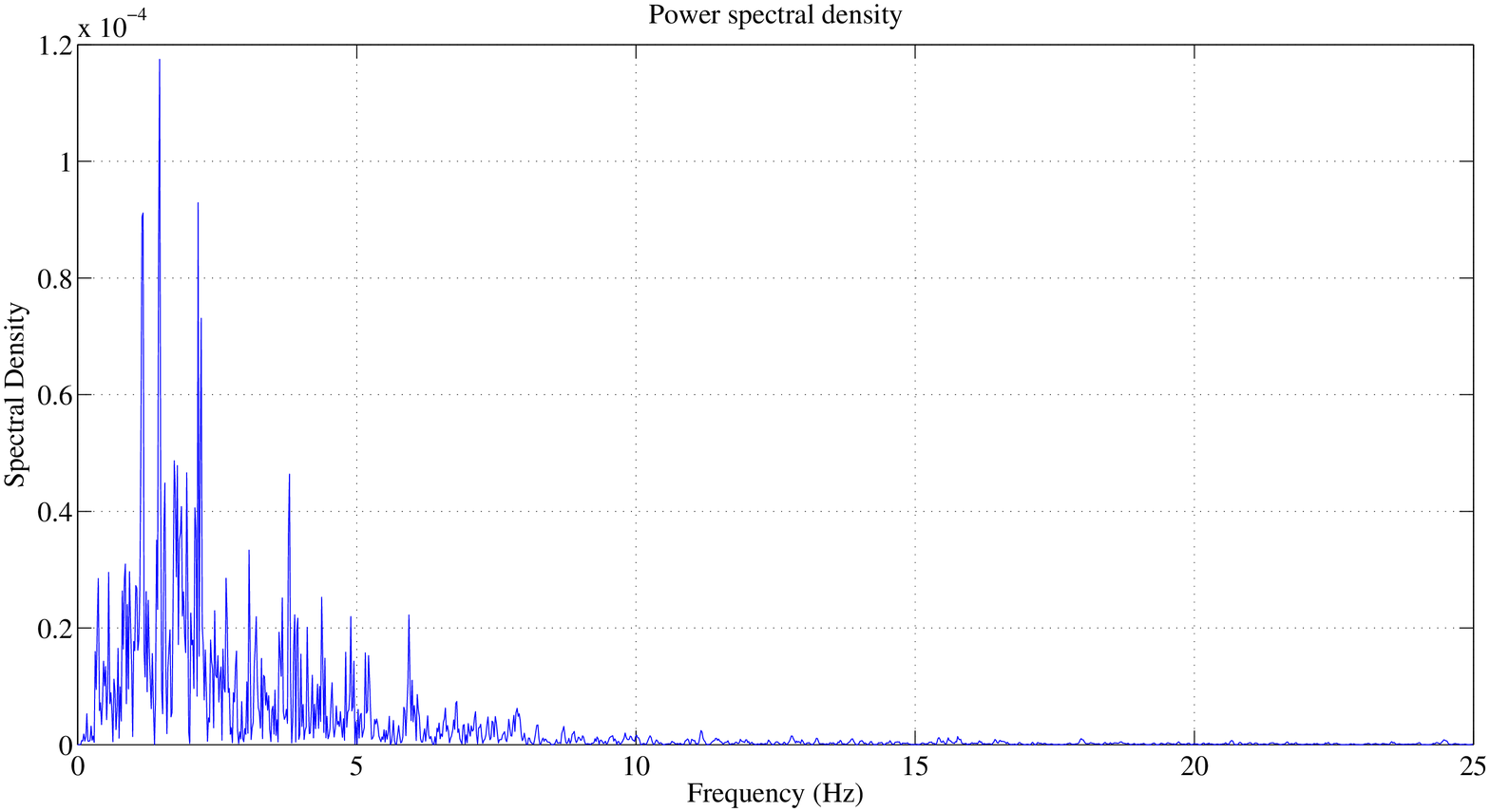}
\includegraphics[angle=0,width=0.5\textwidth,height=0.3\textwidth]{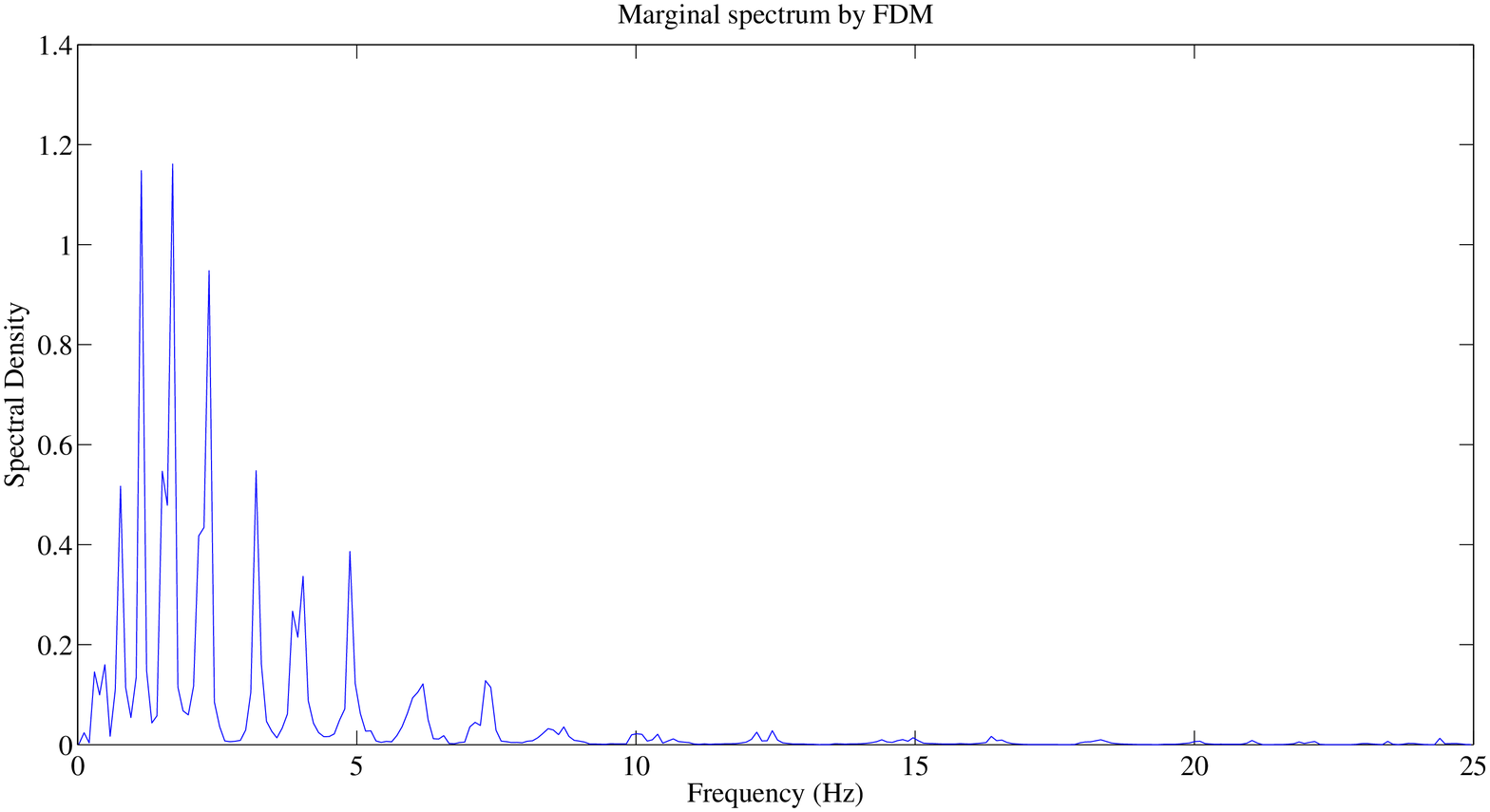}
\includegraphics[angle=0,width=0.5\textwidth,height=0.3\textwidth]{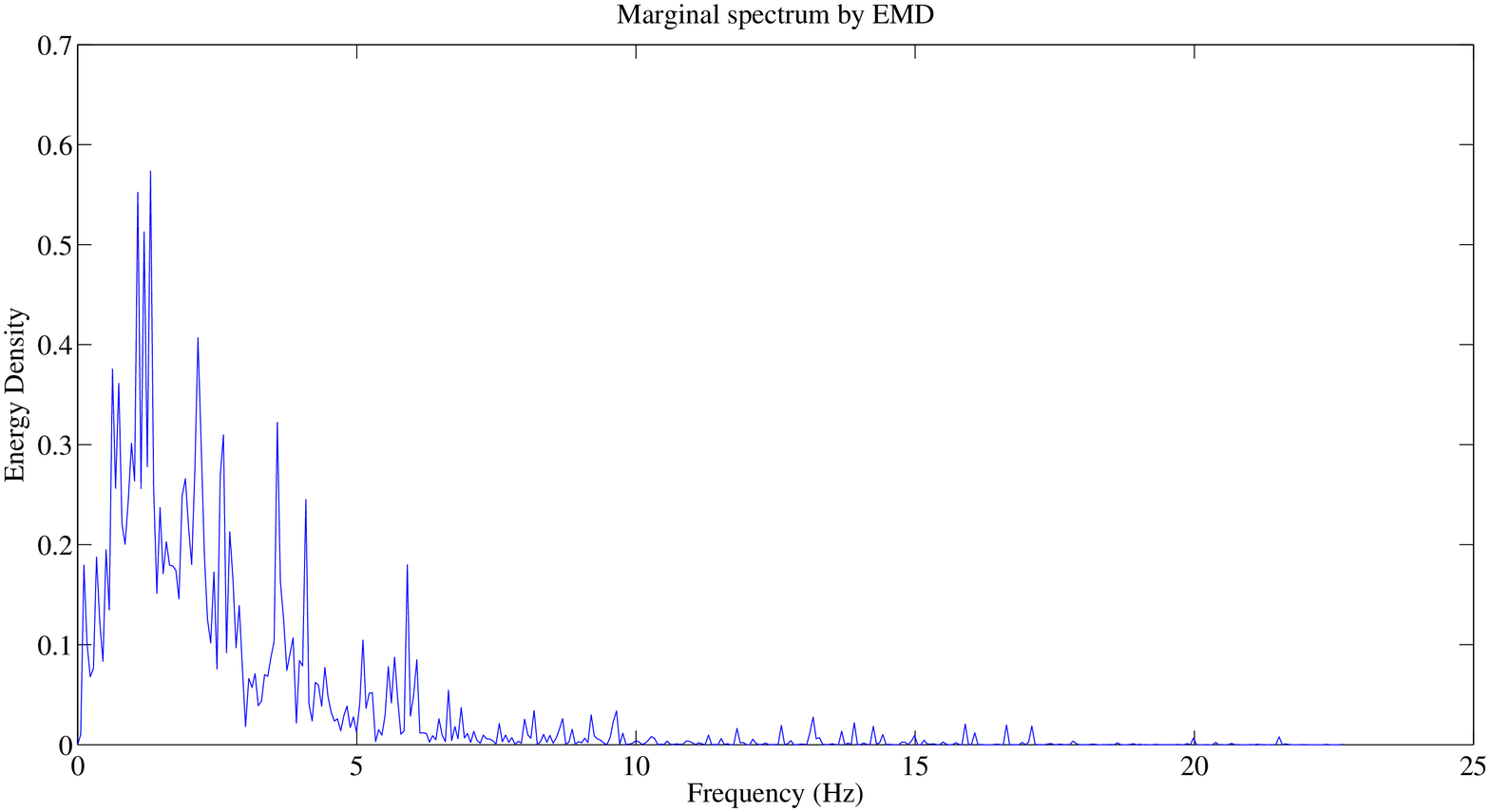}
\captionof{figure}{Plot of (top to bottom): (a) Elcentro Earthquake May 18, 1940 North-South Component, (b) Fourier based power spectral density (PSD), (c) Marginal spectrum by FDM, (d) Marginal spectrum by EMD.}
\label{fig:eq_MS}
\end{figure}
\begin{figure}[!t]
\centering
\includegraphics[angle=0,width=0.5\textwidth,height=0.3\textwidth]{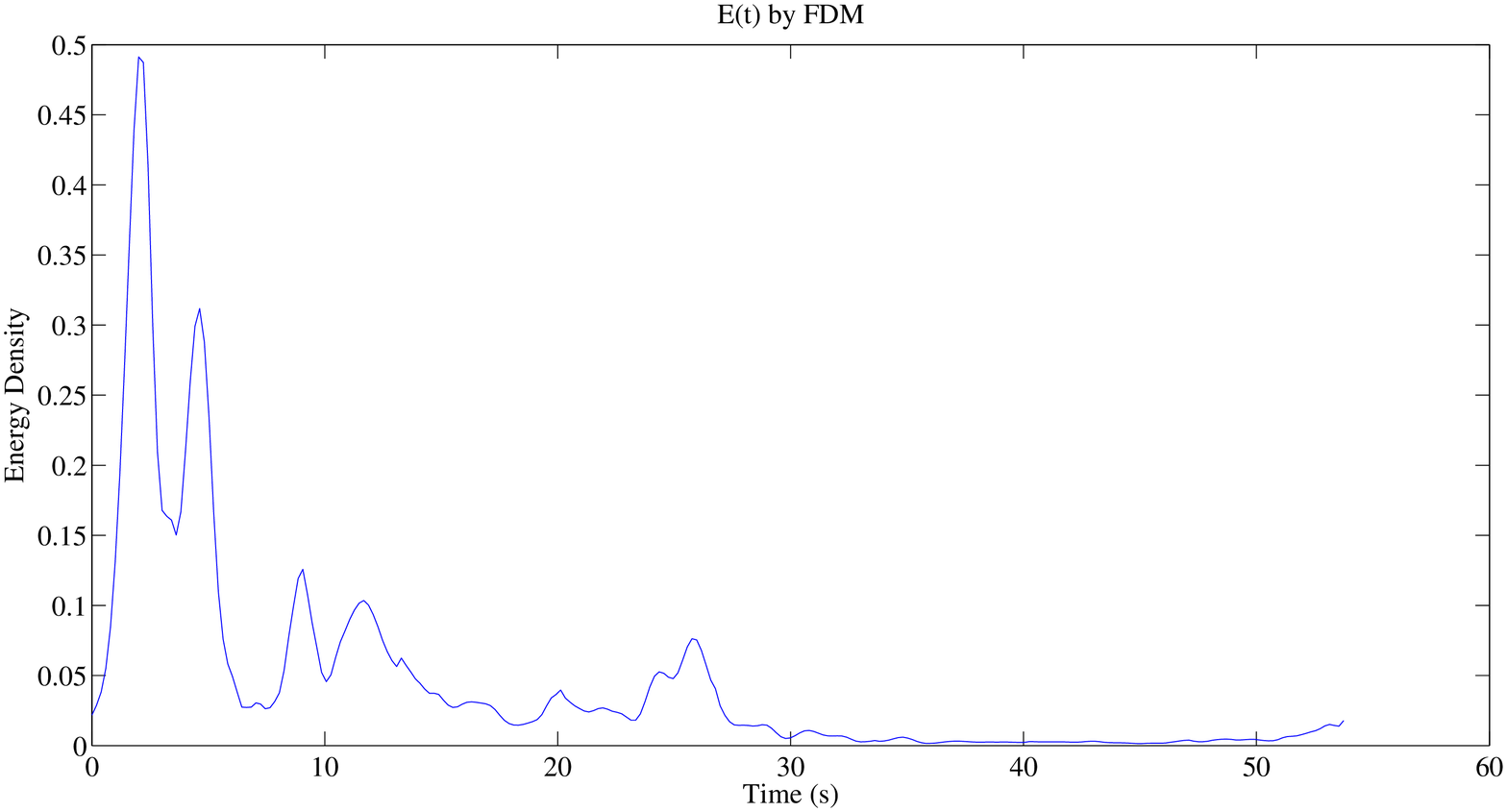}
\includegraphics[angle=0,width=0.5\textwidth,height=0.3\textwidth]{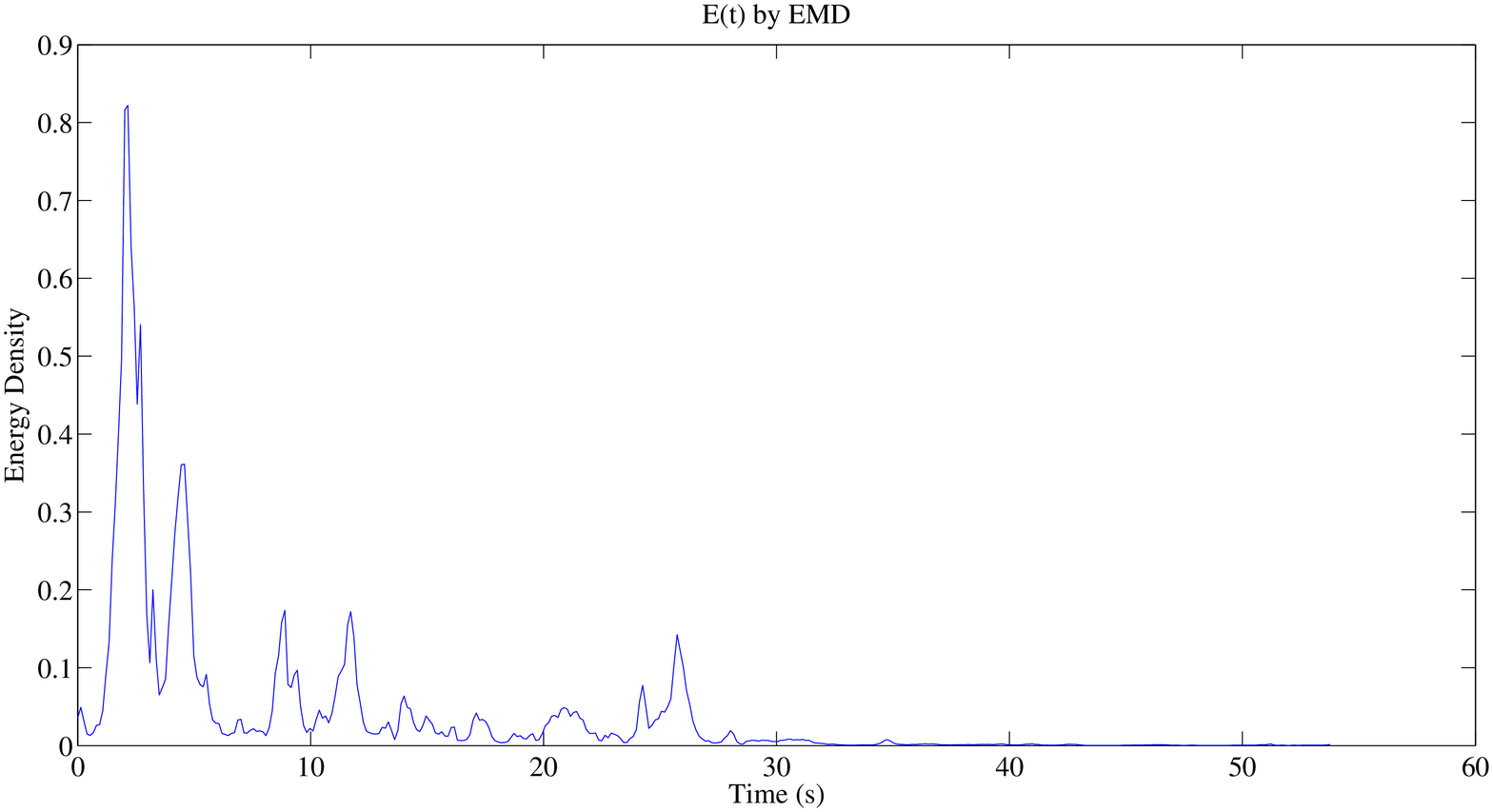}
\captionof{figure}{Plot of the instantaneous energy density $E(t)$ (top to bottom) using the: (a) FDM and (b) EMD.}
\label{fig:eq_Et}
\end{figure}
\begin{figure}[!t]
\centering
\includegraphics[angle=0,width=0.5\textwidth,height=0.3\textwidth]{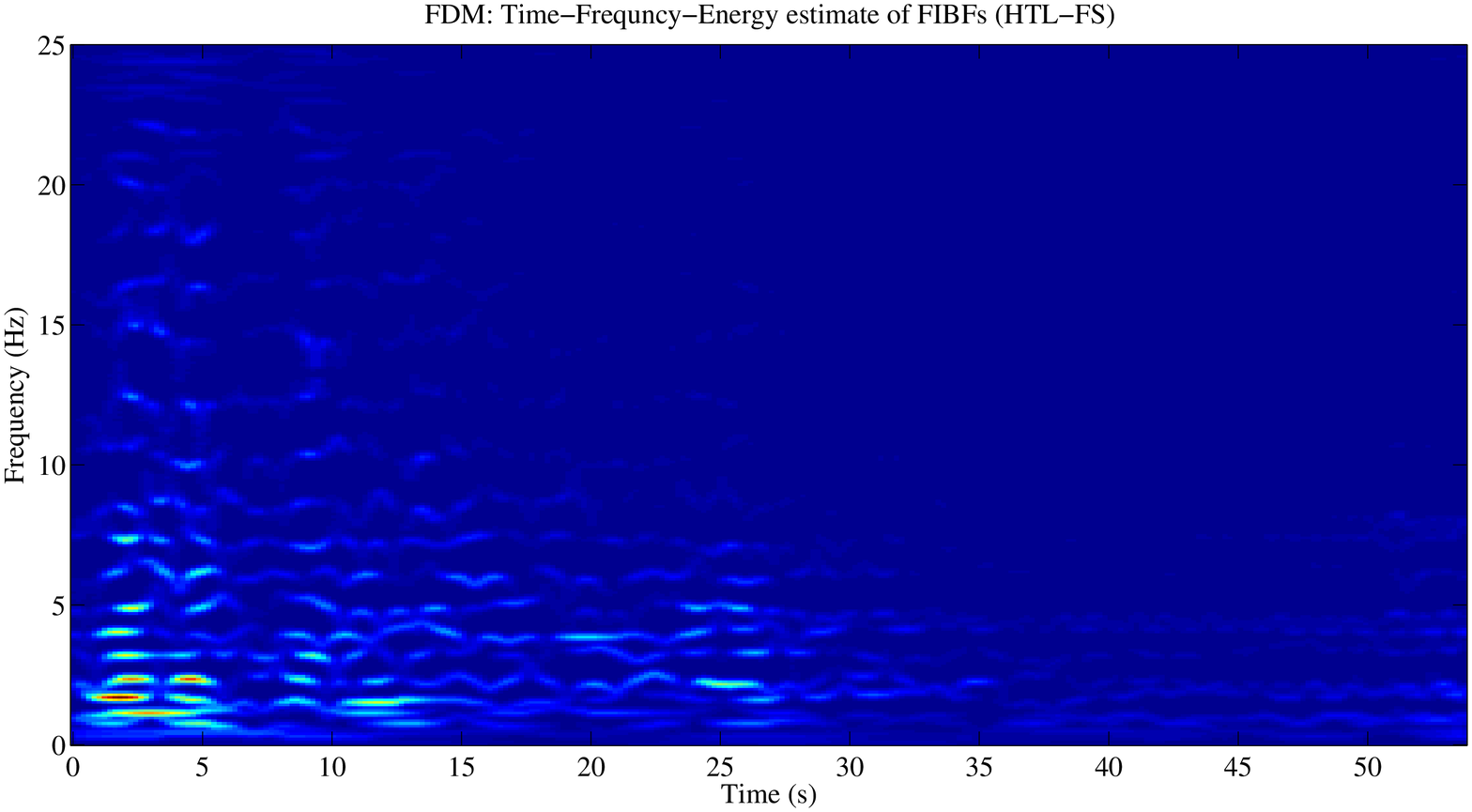}
\includegraphics[angle=0,width=0.5\textwidth,height=0.3\textwidth]{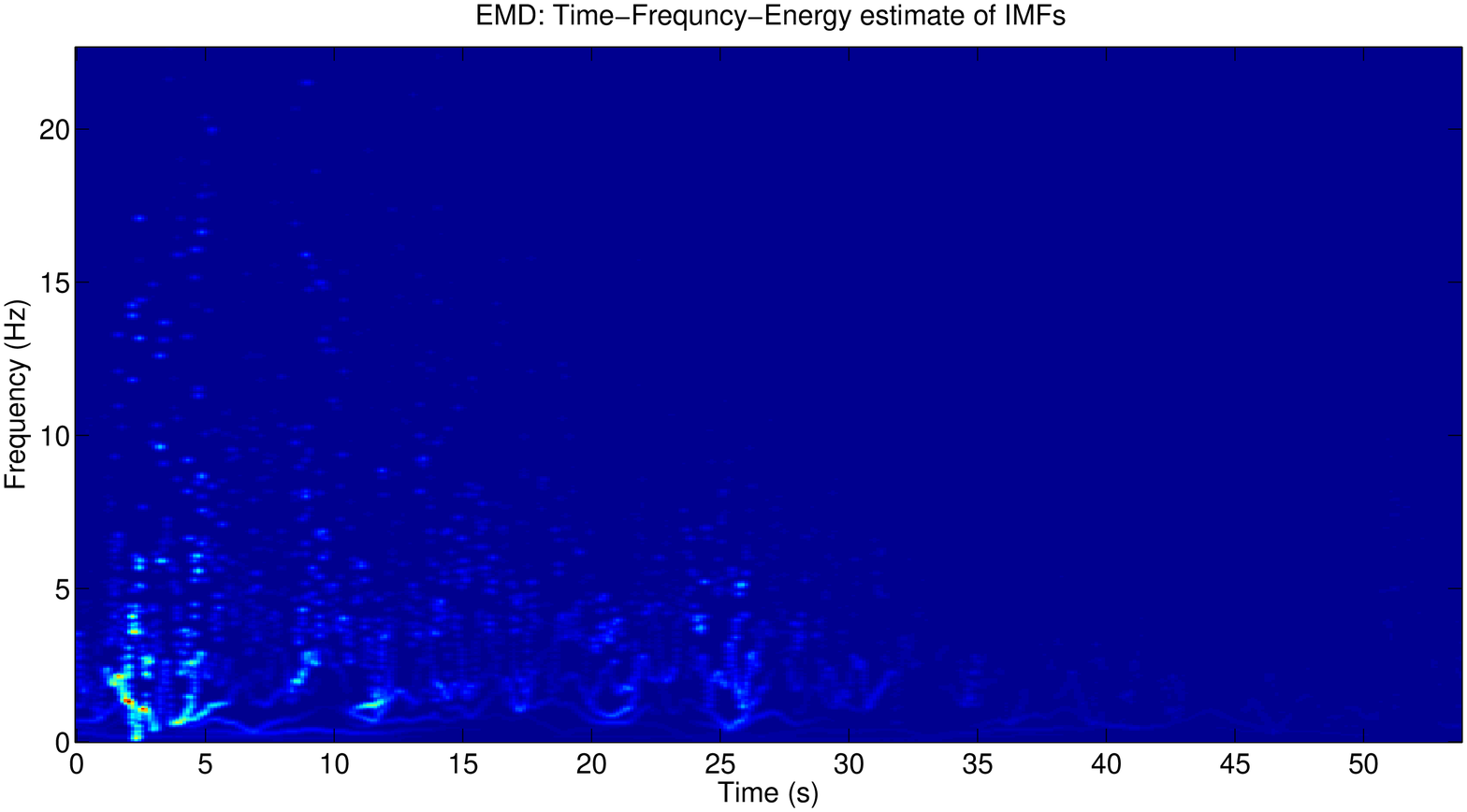}
\includegraphics[angle=0,width=0.5\textwidth,height=0.3\textwidth]{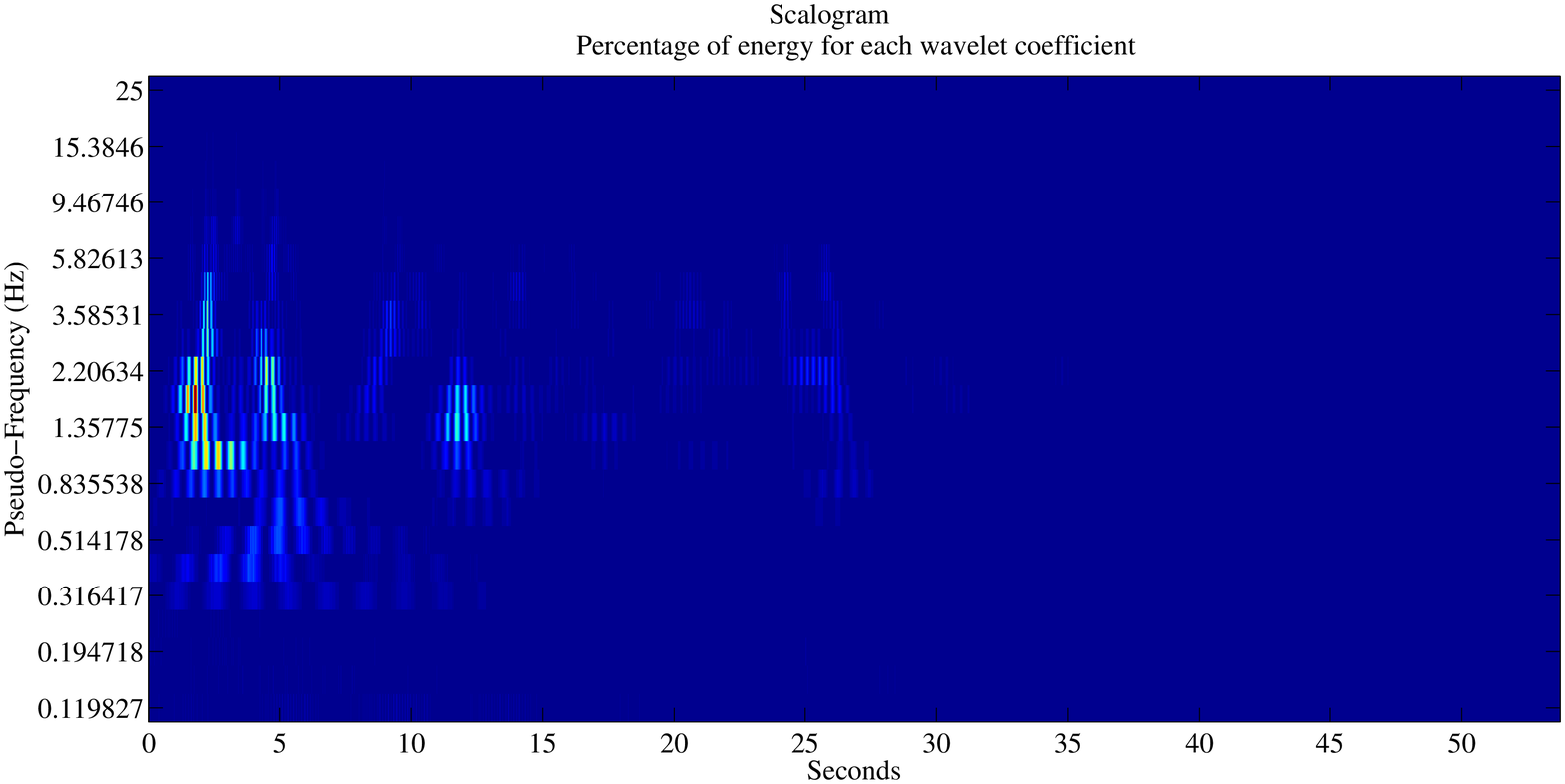}
\captionof{figure}{Plot of TFE (top to bottom) using the: (a) FDM (b) EMD and (c) CWT.}
\label{fig:eq_TFE}
\end{figure}
\noindent Earthquake time series signal is nonlinear and nonstationary data. The Elcentro Earthquake data (sampled at $F_s= 50Hz$) has been taken from~\cite{EQ33} and is shown in Figure~\ref{fig:eq_MS}. The critical frequency range that matter in the structural design is less than $10Hz$, and from the Fourier based power spectral density (PSD), marginal spectrum by FDM and marginal spectrum by EMD in Figure~\ref{fig:eq_MS} show that almost all the energy in this data is within $10Hz$.
The instantaneous energy fluctuations by the FDM and EMD methods, as shown in Figure~\ref{fig:eq_Et}, are similar in nature. The TFE distribution by the FDM, EMD and CWT methods are shown in Figure~\ref{fig:eq_TFE}, and all three methods indicate maximum energy concentration around $1.7Hz$ and 2 second. There is enhanced TFE tracking by FDM methods as it provide better details of how the different waves arrive from the epical center to the recording station, e.g. the compression
waves of small amplitude but higher frequency range of $10$ to $20Hz$, the shear and surface waves of strongest amplitude and lower frequency range of below $5Hz$ which does most of the damage, and other body shear waves which are present over the full duration of the data span.
\subsection{Application to a speech signal analysis}
\begin{figure}[!t]
\centering
\includegraphics[angle=0,width=0.5\textwidth,height=0.3\textwidth]{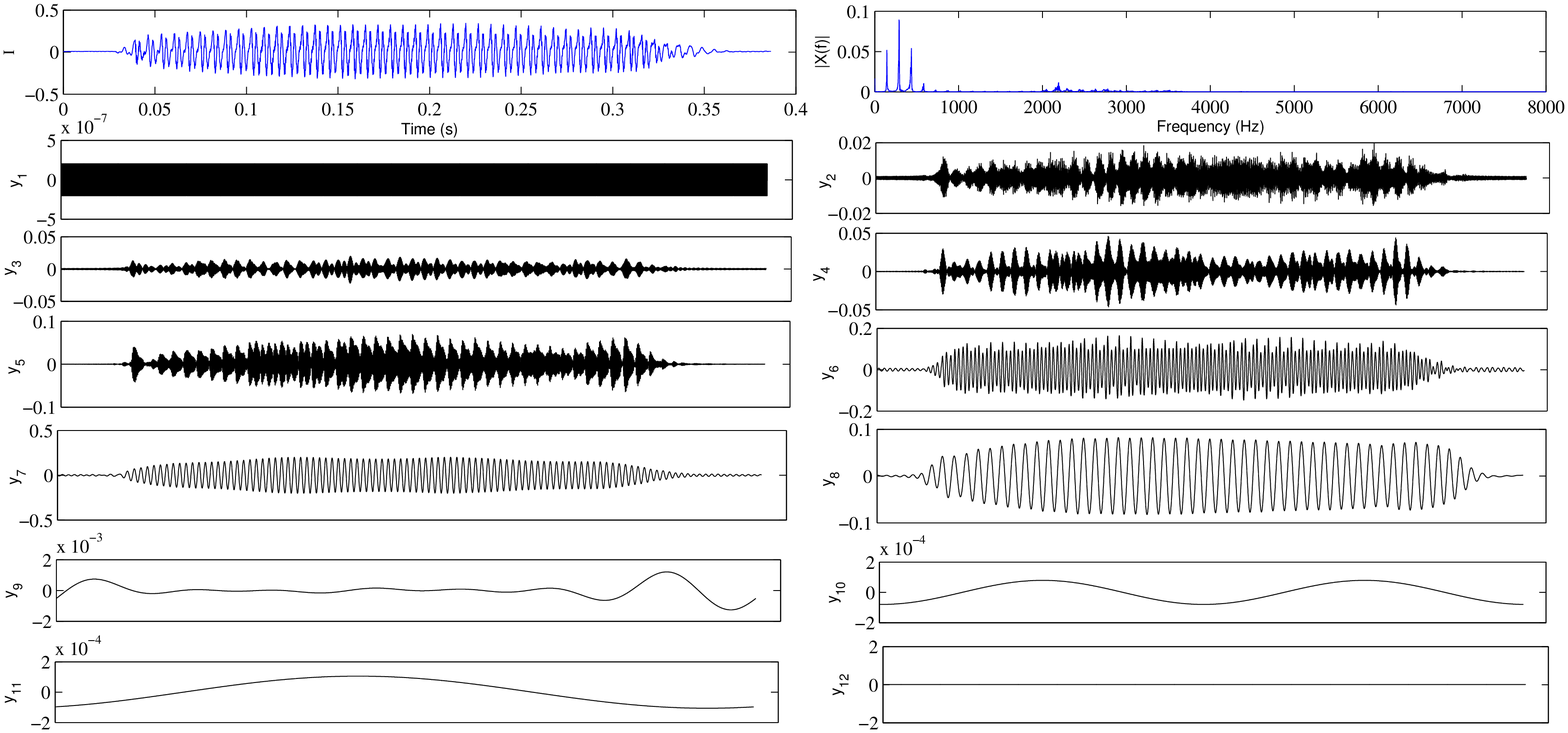}
\captionof{figure}{Vowel `small-cap I' sound with sampling frequency $F_s=16000$ Hz (top left) and its Fourier spectrum (top right). The FDM generates highest frequency component ($y_1$), FIBFs ($y_2$ to $y_{11}$) and DC term ($y_{12}$). The fundamental frequency $F_0$ is captured accurately in FIBF $y_8$.}
\label{fig:speech_FIBFs}
\end{figure}
\begin{figure}[!t]
\centering
\includegraphics[angle=0,width=0.5\textwidth,height=0.3\textwidth]{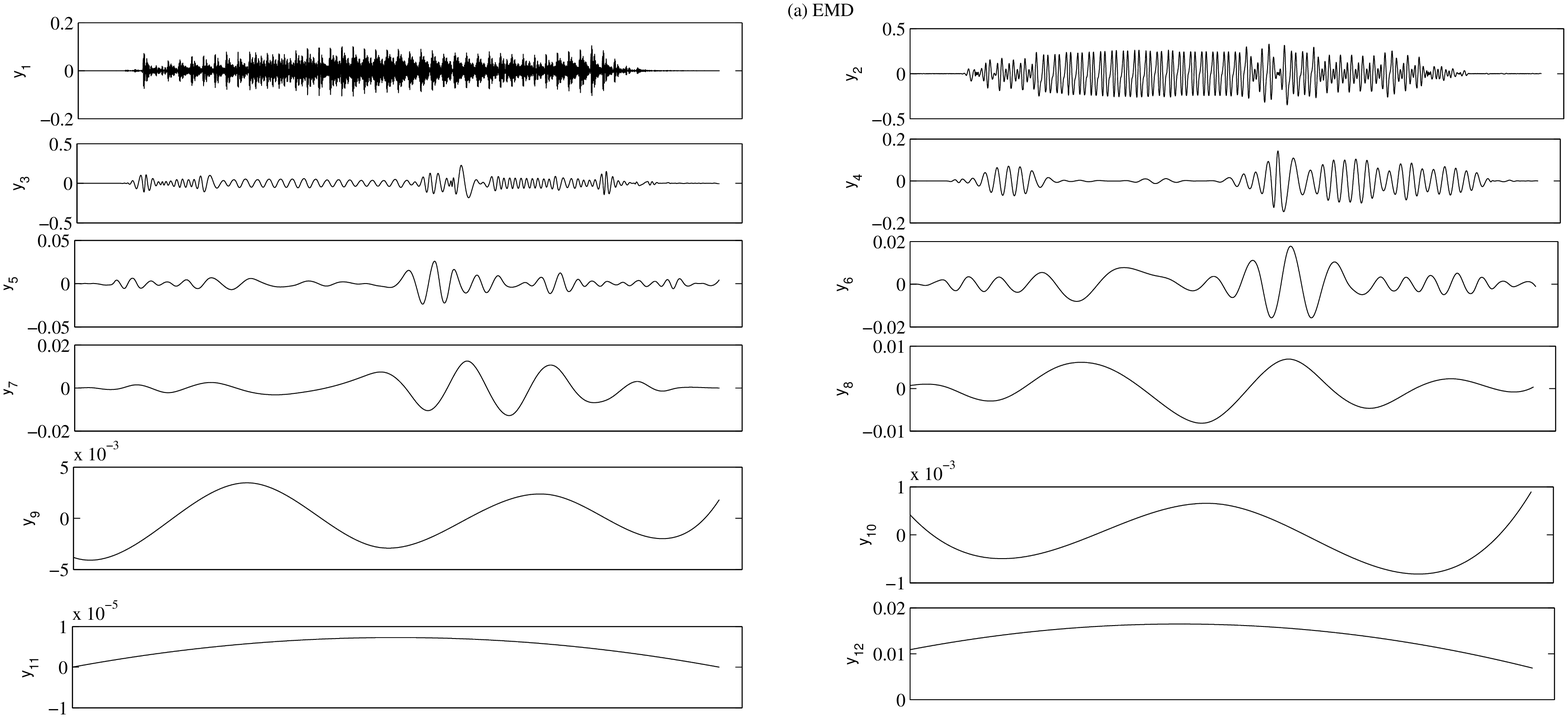}
\includegraphics[angle=0,width=0.5\textwidth,height=0.3\textwidth]{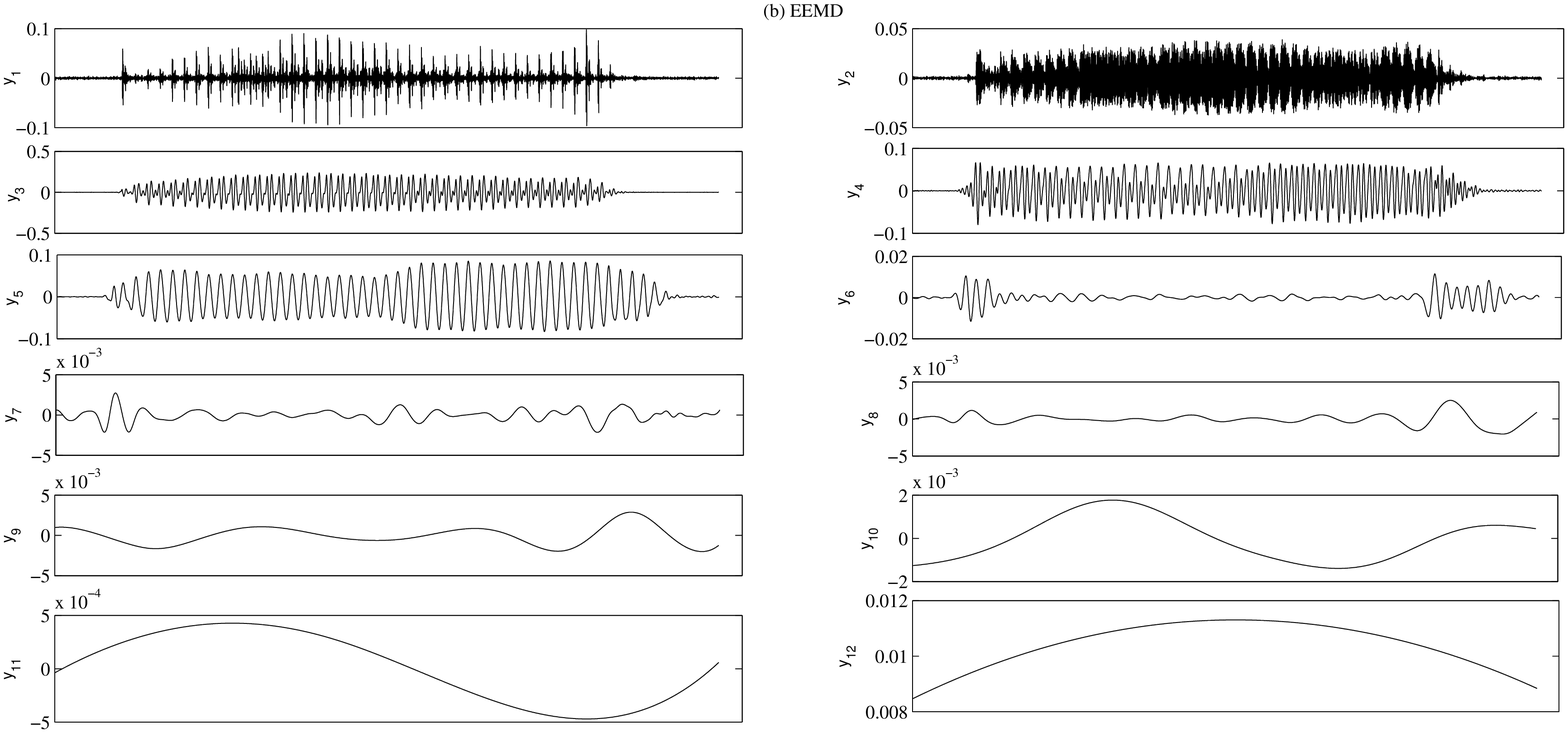}
\captionof{figure}{IMFs ($y_1$ to $y_{12}$) generated by (a) EMD and (b) EEMD (with zero mean and 0.2 standard deviation of the added noise of normal distribution and the ensemble size of 300) algorithms for vowel `small-cap I' with $F_s=16000$ Hz. The EMD is not able to catch $F_0$ in any mode, and the EEMD is able to capture $F_0$ accurately in IMF $y_5$.}
\label{fig:speech_IMFs}
\end{figure}
\begin{figure}[!t]
\centering
\includegraphics[angle=0,width=0.5\textwidth,height=0.3\textwidth]{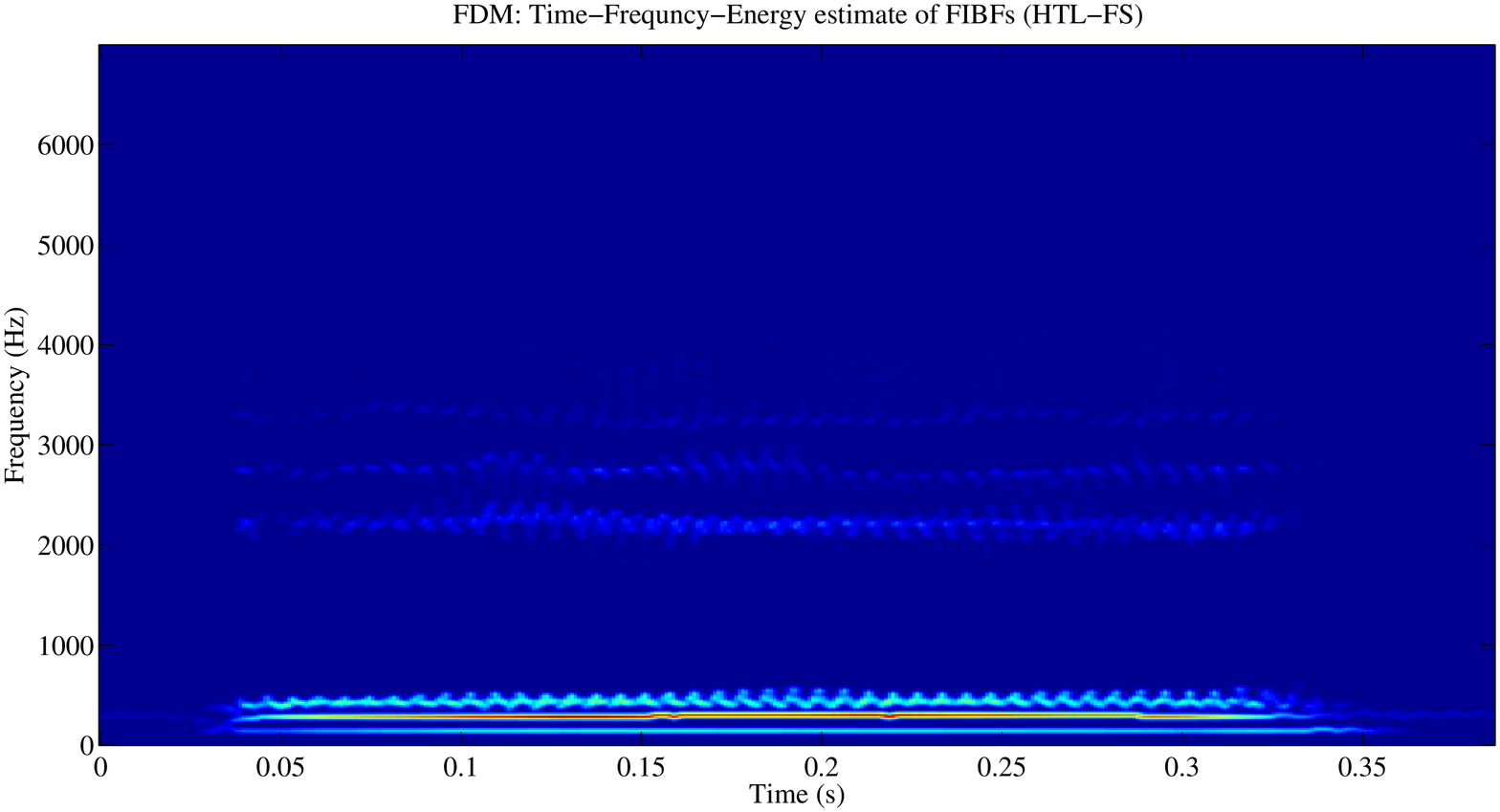}
\includegraphics[angle=0,width=0.5\textwidth,height=0.3\textwidth]{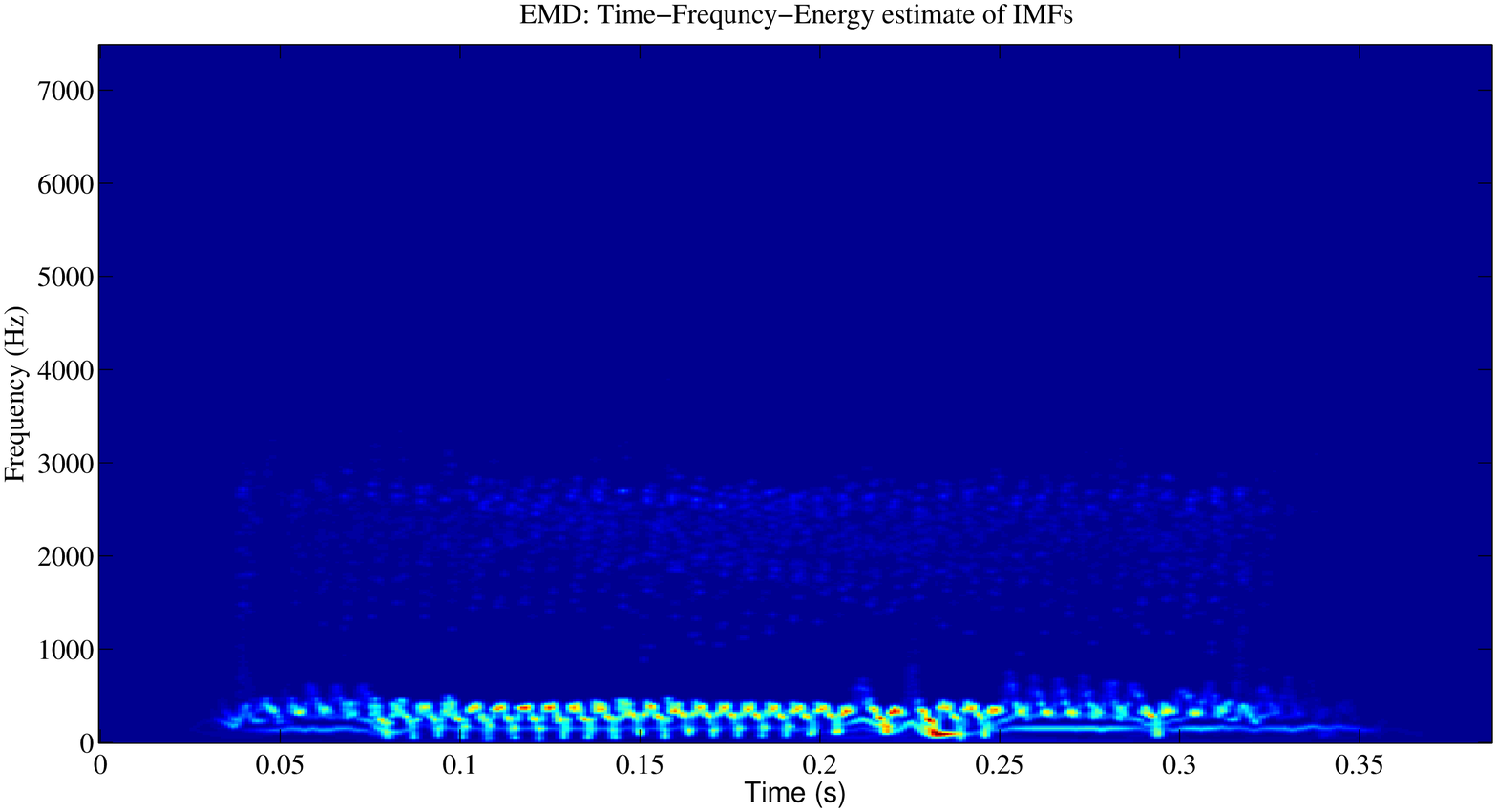}
\includegraphics[angle=0,width=0.5\textwidth,height=0.3\textwidth]{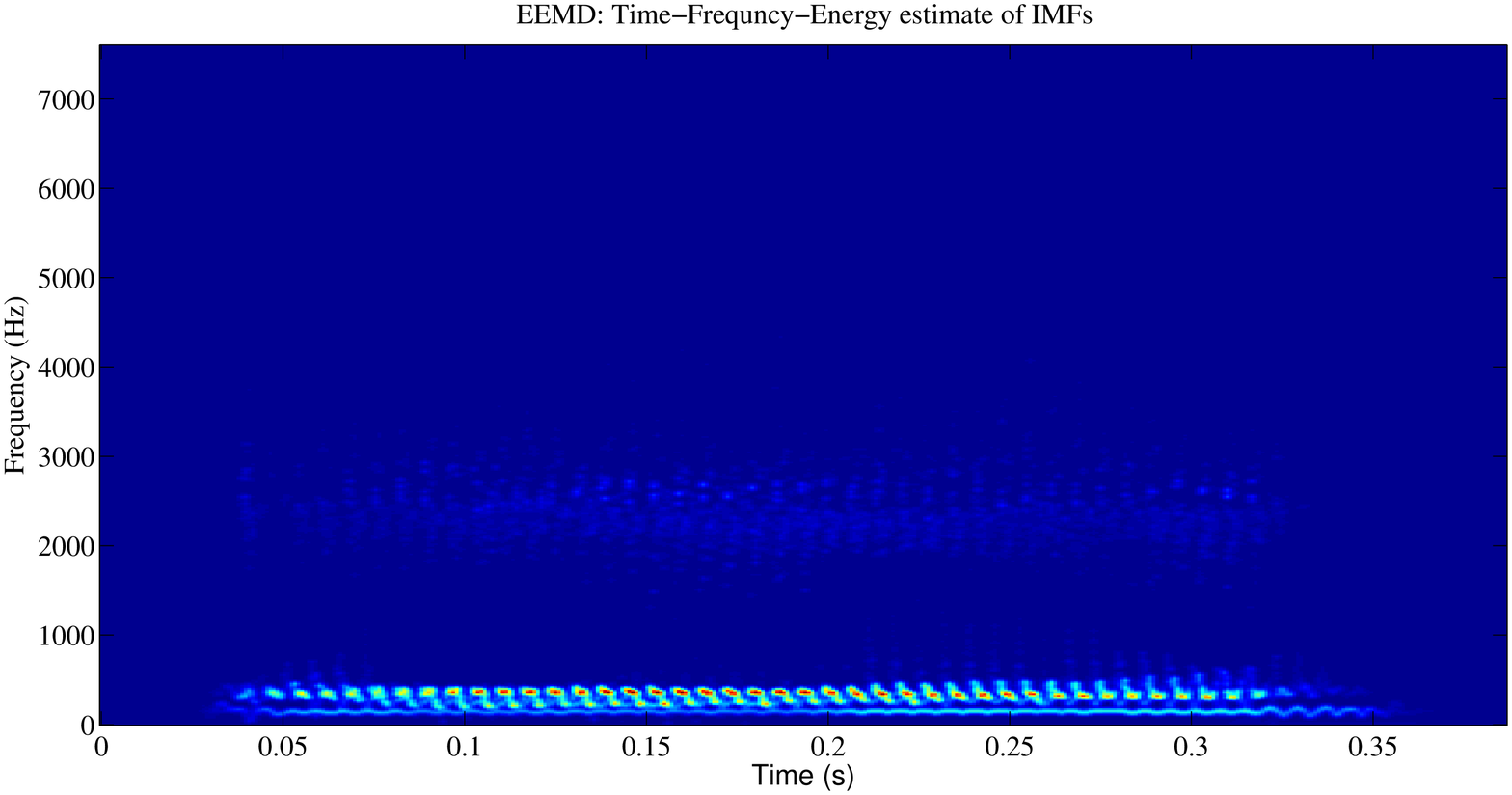}
\captionof{figure}{The TFE plot by FDM (top) EMD (middle) and (b) EEMD of same data (vowel `small-cap I'). There is enhanced TFE tracking when using FDM.}
\label{fig:speech_TFE}
\end{figure}
To illustrate the advantages of FDM in speech signal analysis, we decompose a quasiperiodic voice signal (small-cap I vowel). In Figure~\ref{fig:speech_FIBFs}, the FIBFs $y_2$ to $y_5$ seem to extract most of the noise, while $y_6$ to $y_7$ seem to express most of the signal information at high frequencies. Moreover, the FIBF $y_8$ captures accurately $F0$, the fundamental frequency (inverse of glottal pulse or pitch period length) of the voice signal.
In contrast, EMD (Figure~\ref{fig:speech_IMFs} (a)) presents mode mixing between modes two and three, between modes three and four, and between modes four and five. Also, the EMD is not able to catch $F_0$ in any mode. In Figure~\ref{fig:speech_IMFs} (b), the EEMD's is able to catch $F_0$ in mode $y_5$, although it can be observed that positive minima and negative maxima in the mode $y_4$ and $y_7$, there is clear violation to IMF condition (1). Besides this problem, the major drawbacks in this case are huge computation complexity for this ensemble size and the reconstruction error that may be significant. The computation complexity can be reduced by selecting the smaller ensemble size but that increases the reconstruction error.
Figure~\ref{fig:speech_TFE} shows the plots of TFE obtained by the FDM, EMD and EEMD algorithms for same data (vowel `small-cap I'). Clearly, there is enhanced TFE tracking when using FDM. The results in this application as well are in clear favor of the Fourier method proposed in this paper.
\section{Conclusion}
In this paper, we have proposed: (1) The Fourier Decomposition Method (FDM), for nonlinear and nonstationary time series analysis, which decomposes any data into a small number of `Fourier intrinsic band functions' (FIBFs). The FDM is a generalized Fourier expansion with variable amplitudes and frequencies of a time series by the Fourier method itself. (2) The zero-phase filter bank based multivariate FDM (MFDM) algorithm, for the analysis of multivariate nonlinear and nonstationary time series, which is generating finite number of band limited multivariate FIBFs (MFIBFs). (3) An algorithm to obtain cutoff frequencies required in MFDM algorithm for zero-phase high or low pass filtering of multivariate signals.

The fundamental and conceptual contributions of the this study are the Fourier based decomposition method (i.e. FDM) and the introduction of the FIBFs. The FIBFs form the basis of the decomposition that are complete, orthogonal, local and adaptive. The instantaneous frequencies of the FIBFs produce a time-frequency-energy distribution of any signal. A time-frequency-energy distribution of a signal is used in various fields of science and engineering for analysis of physical phenomena and engineering systems. The proposed methods produce the final presentation of the results in a time-frequency-energy distribution that reveals the imbedded structures of a signal. Unlike the various EMD algorithms, the FDM and MFDM are mathematically well defined, supported by the well established theories of filter and the Fourier transforms. The FDM and MFDM methods do not suffer from mode mixing, detrend uncertainty and end effect artefacts as extraction of FIBFs does not depend on distribution of local extrema across the range of data. Simulation results demonstrate the efficacy of the proposed methods.
\appendix
\subsection{The FDM for continuous time real function}
Let $x(t)$ be a non-periodic, real function of time and follow the Dirichlet conditions, then the Fourier transform (FT) of $x(t)$ is defined as
\begin{equation}
X(f)=\int_{-\infty}^{\infty} x(t) \exp(-j2\pi f t) \ud t \label{FDM_eq1}
\end{equation}
 and inverse Fourier transform is defined as
\begin{equation}
x(t)=\int_{-\infty}^{\infty} X(f) \exp(j2\pi f t) \ud f \label{FDM_eq2}
\end{equation}
 It is easy to show that $\int_{-\infty}^{0} X(f) \exp(j2\pi f t) \ud f= \int_{0}^{\infty} X(-f) \exp(-j2\pi f t) \ud f$. From Eq.~\eqref{FDM_eq1} $X(-f)=X^*(f)$ and, hence, we rewrite Eq.~\eqref{FDM_eq2} as
\begin{equation}
x(t)=\int_{0}^{\infty} [X(f) \exp(j2\pi f t)  +  X^*(f) \exp(-j2\pi f t)] \ud f \label{FDM_eq3}
\end{equation}
In this Eq., second term is complex conjugate of first term. As $x(t)$ is a real function, we can write
\begin{equation}
x(t)=Re\{z(t)\} \label{FDM_eq4}
\end{equation}
where analytic function $z(t)=2\int_{0}^{\infty} X(f) \exp(j2\pi f t) \ud f$ and  $Re\{z(t)\}$ denote real part of function $z(t)$. We write Eq.\eqref{FDM_eq4} as
\begin{equation}
2\int_{0}^{\infty} X(f) \exp(j2\pi f t) \ud f=\sum_{i=1}^M a_i(t)\exp(j\phi_i(t)) \label{FDM_eq5}
\end{equation}
where (with $f_0=0, f_M=\infty$)
\begin{equation}
 a_i(t)\exp(j\phi_i(t)) =  2\int_{f_{i-1}}^{f_i} X(f) \exp(j2\pi f t) \ud f, \label{FDM_eq6}
\end{equation}
for $i=1,\cdots,M$. To obtain minimum number of AFIBFs in low to high frequency scan, for each $i$, start with $f_{i-1}$, increase and select the maximum value of $f_i$ such that $f_{i-1} \le f_i \le \infty$ and
\begin{equation}
a_i(t),f_i(t)=\frac{1}{2\pi}\frac{\ud\phi_i(t)}{\ud t}\ge0, \forall t.\label{FDM_eq6v1}
\end{equation}
Similarly, in high to low frequency scan, the lower and upper limits of integration in~\eqref{FDM_eq6} will change to $f_i$ to $f_{i-1}$, respectively, with
$f_0=\infty, f_M=0$, and we can obtain minimum number of AFIBFs by selecting the minimum value of $f_{i}$ such that $0 \le f_{i} \le f_{i-1}$ and Eq.~\eqref{FDM_eq6v1} is satisfied.

From Eq.~\eqref{FDM_eq3} and~\eqref{FDM_eq4}, it is easy to obtain relationship between the energy of original signal $x(t)$ and energy of its analytic signal $z(t)$ as
\begin{equation}
E_x=\frac{E_z}{2}, \label{FDM_eq6v2}
\end{equation}
that is, the energy of analytic signal is twice of the energy of original signal.

The IF characterizes a local frequency behavior as a function of time. In a dual way, the local time behavior as a function of frequency is characterized by the group delay (GD) : $t_i(f)=-\frac{1}{2\pi}\frac{\ud\phi_i(f)}{\ud f}$. The GD measures the average time arrival of the frequency $f$. In general, the IF and GD define two different curves in the time-frequency plane. Similar to the IF process, for causal signal $x(t)$, we obtain
\begin{equation}
 a_i(f)\exp(-j\phi_i(f)) =  \int_{t_{i-1}}^{t_i} x(t) \exp(-j2\pi f t) \ud t,  \label{FDM_eq8}
\end{equation}
with $t_0=0, t_M=\infty$ such that $a_i(f),t_i(f)=-\frac{1}{2\pi}\frac{\ud\phi_i(f)}{\ud f}\ge0, \forall f$, for $i=1,\cdots,M$.
\subsection{The FDM for discrete time real function}
Let $x[n]$ be a non-periodic, real function of time and follow the Dirichlet conditions, then the discrete time Fourier transform (DTFT) of $x[n]$ is defined as
\begin{equation}
X(\omega)=\sum_{-\infty}^{\infty} x[n] \exp(-j\omega n)  \label{FDM_eq21}
\end{equation}
 and inverse discrete time Fourier transform (IDTFT) is defined as
\begin{equation}
x[n]=\frac{1}{2\pi}\int_{-\pi}^{\pi} X(\omega) \exp(j\omega n) \ud \omega \label{FDM_eq22}
\end{equation}
 It is easy to show that $\int_{-\pi}^{0} X(\omega) \exp(j\omega n) \ud \omega = \int_{0}^{\pi} X(-\omega) \exp(-j\omega n) \ud \omega$. From Eq.~\eqref{FDM_eq21} $X(-\omega)=X^*(\omega)$ and, hence, we rewrite Eq.~\eqref{FDM_eq22} as
\begin{equation}
x[n]=\frac{1}{2\pi}[\int_{0}^{\pi} X(\omega) \exp(j\omega n) \ud \omega +\int_{0}^{\pi} X^*(\omega) \exp(-j\omega n) \ud \omega]. \label{FDM_eq23}
\end{equation}
In this Eq., second term is complex conjugate of first term. As $x[n]$ is real function, we can write
\begin{equation}
x[n]=Re\{z[n]\} \label{FDM_eq24}
\end{equation}
where analytic signal $z[n]=\frac{1}{\pi}\int_{0}^{\pi} X(\omega) \exp(j\omega n) \ud \omega$ and $Re\{z[n]\}$ denote real part of function $z[n]$. We write Eq.~\eqref{FDM_eq24} as
\begin{equation}
\frac{1}{\pi}\int_{0}^{\pi} X(\omega) \exp(j\omega n) \ud \omega=\sum_{i=1}^M a_i[n]\exp(j\phi_i[n]) \label{FDM_eq25}
\end{equation}
where (with $\omega_0=0, \omega_M=\pi$)
\begin{equation}
 a_i[n]\exp(j\phi_i[n]) =  \frac{1}{\pi}\int_{\omega_{i-1}}^{\omega_i} X(\omega) \exp(j\omega n) \ud \omega, \label{FDM_eq26}
\end{equation}
for $i=1,\cdots,M$.
In order to obtain minimum number of AFIBFs in low to high frequency scan, for each $i$, start with $\omega_{i-1}$, increase and select the maximum value of $\omega_i$ such that $\omega_{i-1} \le \omega_i \le \pi$ and phase $\phi_i[n]$ is a monotonically increasing function, i.e.
\begin{equation}
a_i[n],\omega_i[n]=(\phi_i[n+1]-\phi_i[n])\ge0, \forall n. \label{FDM_eq26v1}
\end{equation}
Similarly, in high to low frequency scan, the lower and upper limits of integration in Eq.~\eqref{FDM_eq26} will change to $\omega_i$ to $\omega_{i-1}$, respectively, with $\omega_0=\pi, \omega_M=0$, and we can obtain minimum number of AFIBFs by selecting the minimum value of $\omega_{i}$ such that $0 \le \omega_{i}\le \omega_{i-1}$ and Eq.~\eqref{FDM_eq26v1} is satisfied.

\section*{Acknowledgment}
The authors would like to thank JIIT Noida for permitting to carry out research at IIT Delhi.
\ifCLASSOPTIONcaptionsoff
  \newpage
\fi


\begin{thebibliography}{1}
\bibitem{rs1} {Huang N.E., Shen Z., Long S., Wu M., Shih H., Zheng Q., Yen N., Tung C., and Liu H.}, {``The empirical mode decomposition and Hilbert spectrum for non-linear and non-stationary time series analysis}," \emph{Proc. R. Soc. A}, (1988) 454, 903--995.
\bibitem{rs25} {Costa M., Priplata A.A., Lipsitz L.A., Wu Z., Huang N.E., Goldberger A.L., and Peng C.K.}, {``Noise and poise: enhancement of postural complexity in the elderly with a stochastic-resonance-based therapy}," \emph{Europhys. Lett.}, (2007) EPL 77, 68008.
\bibitem{rs26} {Cummings D.A., Irizarry R.A., Huang N.E., Endy T.P., Nisalak A., Ungchusak K., and Burke D.S.}, {``Travelling waves in the occurrence of dengue haemorrhagic fever in Thailand}," \emph{Nature}, (2004) 427, 344--347.
\bibitem{rs30} {Hu K., Peng C.K., Huang N.E., Wu Z., Lipsitz L.A., Cavallerano J., and Novaka V.}, ``Altered phase interactions between spontaneous blood pressure and flow fluctuations in type 2 diabetes mellitus: nonlinear assessment of cerebral autoregulation," \emph{Physica A} (2008) 387 (10), 2279--2292.
\bibitem{rs31} {Lo M.T., Lin L.Y., Hsieh W.H., Ko P.C.I., Liu Y.B., Lin C., Chang Y.C., Wang C.Y., Young V.H.W., Chiang W.C., Lin J.L., Chen W.J., and Ma M.H.M.}, ``A new method to estimate the amplitude spectrum analysis of ventricular fibrillation during cardiopulmonary resuscitation," \emph{Resuscitation}, (2013) 84 (11), 1505--1511.
\bibitem{rs27} {Huang N.E., and Wu Z.}, {``A review on Hilbert--Huang transform: Method and its applications to geophysical studies}," \emph{Rev. Geophys.}, (2008) 46, RG2006.
\bibitem{rs29} {Wu Z., Norden E.H., and Chen X.}, ``The multi-dimensional ensemble empirical mode decomposition method," \emph{Adv. Adapt. Data Anal.} (2009) 1, 339--372.
\bibitem{rs4} {Wu Z. and Huang N.E.}, {``Ensemble Empirical Mode Decomposition: a noise-assisted data analysis method}," \emph{Advances in Adaptive Data Analysis}, (2009) 1 (1), 1--41.
\bibitem{rs5} {Rehman N. and Mandic D. P.}, {``Multivariate empirical mode decomposition}," \emph{Proc. R. Soc. A}, (2010) 466, 1291--1302.
\bibitem{rs12} {Chu P. C., Fan C., and Huang N.}, {``Compact Empirical Mode Decomposition: An Algorithm to reduce mode mixing, end effect, and detrend uncertainty}," \emph{Adv. Adapt. Data Analys}, (2012) 4 (3), 1250017.
\bibitem{rs13} {Singh P., Srivastavay P.K., Patney R.K., Joshi S.D. and Saha K.}, {``Nonpolynomial Spline Based Empirical Mode Decomposition}," \emph{2013 International Conference on Signal Processing and Communication}, (2013) 435-440.
 \bibitem{rs14} {Singh P., Patney R.K., Joshi S.D. and Saha K.}, {``Some studies on nonpolynomial interpolation and error analysis}," \emph{Applied Mathematics and Computation}, (2014) 244, 809--821.
\bibitem{rs19} { Singh P., Patney R.K., Joshi S.D. and Saha K.}, {``The Hilbert spectrum and the Energy Preserving Empirical Mode Decomposition},"  \emph{arXiv:1504.04104v1 [cs.IT]}.
\bibitem{rs22} {Mandic D. P., Rehman N., Wu Z. and Huang N.E.}, {``Empirical Mode Decomposition-Based Time-Frequency Analysis of Multivariate Signals}," \emph{IEEE signal processing magazine}, (2013) November, 74--86.
\bibitem{rs15} {Flandrin P., Rilling G., and Goncalves P.}, {``Empirical mode decomposition as a filter bank}," \emph{IEEE Signal Process. Lett.}, (2004) 11 (2), 112--114.
\bibitem{rs16} {Rehman N. ur and Mandic D.P.}, {``Filter Bank Property of Multivariate Empirical Mode Decomposition}," \emph{IEEE Transactions on Signal Processing}, (2011) 59 (5), 2421--2426.
\bibitem{rs28} {Priestley M. B.}, ``Evolutionary spectra and non-stationary processes," \emph{J. R. Statist. Soc.}, (1965) B27, 204--237.
\bibitem{rs17} {Daubechies I., Lu J., and Wu H.}, {``Synchrosqueezed wavelet transforms: An empirical mode decomposition-like tool}," \emph{Applied and Computational Harmonic Analysis}, (2011) 30, 243--261.
\bibitem{rs20} {Bedrosian E.}, {``A product threom for Hilbert transform}," \emph{Proc. of IEEE}, (1963) 51 (5), 868--869.
\bibitem{rs21} {Nuttall A. H., and Bedrosian E. }, {``On the quadrature approximation to the Hilbert transform of modulated signals}," \emph{Proc. of IEEE}, (1966) 54 (10), 1458--1459.
\bibitem{rs23} Available: http://www.commsp.ee.ic.ac.uk/$\sim$mandic/research/emd.htm
\bibitem{rs24} Available: http://rcada.ncu.edu.tw/research1.htm
\bibitem{rs32} {Thomas Y. H., and Zuoqiang S.}, {``Adaptive data analysis via sparse time-frequency representation}," \emph{Advances in Adaptive Data Analysis}, (2011) 3 (1 \& 2), 1--28.
\bibitem{th4} {Boashash B.}, ``Estimating and interpreting the instantaneous frequency of a signal. I. Fundamentals," \emph{Proc. IEEE}, (1992) 80 (4), 520--538.
\bibitem{EQ33} [Online]. Available: http://www.vibrationdata.com/elcentro.htm
\bibitem{ucrtp} {Huang N. E., Chen X., Lo M. T. and Wu Z.}, ``On Hilbert Spectral Representation: a True Time-Frequency Representation for nonlinear and nonstationary Data," \emph{Adv. Adapt. Data Anal.} (2011) 63 (03), 63--93.
\bibitem{CompFHT} {Huang N. E., and Wu Z.}, ``A review on Hilbert-Huang transform: Method and its applications to geophysical studies," \emph{Reviews of Geophysics}, (2008) 46 (02), 1--23.

\end{thebibliography}
\end{document}